\documentclass[prd,twocolumn,showpacs,preprintnumbers,amsmath,amssymb,superscriptaddress,floatfix]{revtex4}

\usepackage{graphicx}
\usepackage{dcolumn}
\usepackage{bm}
\usepackage{latexsym}
\newcommand{\beqn}{\begin{eqnarray}}
\newcommand{\eeqn}{\end{eqnarray}}
\newcommand{\eq}[1]{(\ref{#1})}
\newcommand{\cC}{{\cal C}}
\newcommand{\cL}{{\cal L}}
\newcommand{\cZ}{{\cal Z}}
\newcommand{\dD}{{\mathrm D}}
\newcommand{\dd}{{\mathrm d}}

\begin{document}

\preprint{KANAZAWA/04-15}
\preprint{ITEP-LAT/2005-14}

\title{Vacuum type of $\mathbf{SU(2)}$ gluodynamics in  maximally Abelian and Landau gauges}
\author{M. N. Chernodub}\affiliation{Institute of Theoretical and Experimental Physics ITEP, 117259 Moscow, Russia}
\author{Katsuya Ishiguro}
\affiliation{Institute for Theoretical Physics, Kanazawa University,
Kanazawa 920-1192, Japan}
\affiliation{RIKEN, Radiation Laboratory, Wako 351-0158, Japan}
\author{Yoshihiro Mori}
\affiliation{Institute for Theoretical Physics, Kanazawa University,
Kanazawa 920-1192, Japan}
\affiliation{RIKEN, Radiation Laboratory, Wako 351-0158, Japan}
\author{Yoshifumi Nakamura}
\affiliation{Institute for Theoretical Physics, Kanazawa University,
Kanazawa 920-1192, Japan}
\affiliation{RIKEN, Radiation Laboratory, Wako 351-0158, Japan}
\author{M. I. Polikarpov}\affiliation{Institute of Theoretical and Experimental Physics ITEP, 117259 Moscow, Russia}
\author{Toru Sekido}
\affiliation{Institute for Theoretical Physics, Kanazawa University,
Kanazawa 920-1192, Japan}
\affiliation{RIKEN, Radiation Laboratory, Wako 351-0158, Japan}
\author{Tsuneo Suzuki}
\affiliation{Institute for Theoretical Physics, Kanazawa University,
Kanazawa 920-1192, Japan}
\affiliation{RIKEN, Radiation Laboratory, Wako 351-0158, Japan}
\author{V. I. Zakharov}
\affiliation{Max-Planck Institut f\"ur Physik, F\"ohringer Ring 6, 80805,
M\"unchen, Germany
}

\date{2 August, 2005}

\begin{abstract}
The vacuum type of  $SU(2)$ gluodynamics is studied using
Monte-Carlo simulations in maximally Abelian (MA) gauge and in
Landau (LA) gauge, where the dual Meissner effect is observed to
work. The dual Meissner effect is characterized by the coherence
and the penetration lengths. Correlations between Wilson loops and
electric fields are evaluated in order to measure the penetration
length in both gauges. The coherence length is shown to be fixed
in the MA gauge from measurements of the monopole density around
the static quark-antiquark pair. It is also shown numerically that
a dimension 2 gluon operator $A^+A^-(s)$ and the monopole density
has a strong correlation as suggested theoretically. Such a
correlation is observed also between the monopole density and
$A^2(s)= A^+A^-(s) + A^3A^3(s)$ condensate if the remaining $U(1)$
gauge degree of freedom is fixed to $U(1)$ Landau gauge (U1LA).
The coherence length is determined numerically also from correlations
between Wilson loops and $A^+A^-(s)$ and $A^2(s)$ in MA + U1LA
gauge. Assuming that the same physics works in the LA gauge, we
determine the coherence length from correlations between Wilson
loops and $A^2(s)$. Penetration lengths and coherence lengths in
the two gauges are almost the same. The vacuum type of the
confinement phase in both gauges is near to the border between the
type 1 and the type 2 dual superconductors.
\end{abstract}


\maketitle

\section{\label{sec:level1} Introduction}
It is conjectured that the dual Meissner effect is the color
confinement mechanism~\cite{tHooft:1975pu, Mandelstam:1974pi}. The
conjecture seems to be realized if we perform Abelian
projection~\cite{tHooft:1981ht} in the maximally Abelian (MA) gauge~\cite{suzuki-83,kronfeld}.
Abelian component of the gluon field and Abelian monopoles are found to be
dominant~\cite{AbelianDominance,Reviews}. Abelian electric field is
squeezed by solenoidal monopole currents~\cite{bali-96,Koma:2003gq,Singh:1993jj}.
Monopole condensation is confirmed by the energy-entropy balance
of the monopole trajectories~\cite{shiba95} and by evaluation of the monopole creation
operator~\cite{ref:order:parameter}.
All these facts support the conjecture that the color
confinement is due to the dual Meissner effect caused by the monopole
condensation. Numerical calculations show that the vacuum of quenched
$SU(2)$ QCD ($SU(2)$ gluodynamics) is near the border between the type
1 and the type 2 dual superconductor~\cite{Cea:1995zt,Singh:1993jj,matsubara94,kato98},
although there are some claims that it is a superconductor of type 1,
Ref.~\cite{bali98,Koma:2003gq}. Since the definition of
a dual Higgs field is unknown the coherence length was calculated using classical
Ginzburg-Landau equations, while the penetration length can be
calculated directly measuring the correlations between Wilson loops and
Abelian or non-Abelian electric fields.

In this note, we show that the coherence length could be derived
directly from the measurement of the monopole density around a chromomagnetic flux spanned
between a static quark-antiquark pair. We use the dual Ginzburg-Landau
effective theory of infrared $SU(2)$ gluodynamics~\cite{suzuki88,maedan89}, evaluate the monopole density around the
flux theoretically and compare it with the value obtained numerically.

We consider also the dimension 2 gluon operator
$$
A^+A^-(s)\equiv\sum_{\mu}[(A^1_{\mu}(s))^2+(A^2_{\mu}(s))^2]
$$
in the MA gauge. The MA gauge is a gauge which minimizes a functional
$\sum_{s}A^+A^-(s)$. It is well known that the off-diagonal gluon fields
$A^i_{\mu}(s)$ with $i=1,2$ are small everywhere except at the
sites where monopoles exist. Hence a strong correlation between
$A^+A^-(s)$ and  monopole currents $|k_{\mu}(s)|$ is expected. The
off-diagonal gluons have no essential effects on the confinement of
fundamental charge, whereas they can explain the screening of adjoint
charge~\cite{screening}. If we perform the additional U1LA gauge
fixing after the MA gauge is fixed, the operator $\sum_{s,\mu}(A^{3}_{\mu}(s))^2)$
can have a physical meaning. It is expected from the previous results suggesting
monopole dominance\cite{Reviews} that monopole contribution could
explain all non-perturbative effects in the quantity
$$
A^3A^3(s)\equiv\sum_{\mu}(A^3_{\mu}(s))^2\,.
$$
Hence we expect that the
coherence length can also be derived from correlations between the Wilson
loops and the dimension 2 operator $A^+A^-(s)$ or between the Wilson loops
and the dimension 2 operator
$$
A^2(s)=A^+A^-(s)+A^3A^3(s)\,.
$$
We find
that the coherence lengths determined from the monopole density and the
dimension 2 operators are consistent with each other.
We also observe that the penetration length and the coherence length are almost the
same and we conclude that the vacuum is near the border between the type 1 and
the type 2 dual superconductors in the MA gauge.

Since the MA gauge -- in which the confinement mechanism is definitely found to be realized --
is just one gauge among infinite possible gauges and, on the other hand, the physics should be
gauge-independent it is important to know the confinement mechanism as well as the type of the
vacuum in another gauge and in the case in which the gauge fixing is not performed at all.

This problem has been discussed recently in
Ref.~\cite{suzuki05} where the Landau (LA) gauge is considered and for
Abelian components the dual Meissner effect is observed. A magnetic
displacement current plays the role of the solenoidal super current
which squeezes the Abelian electric fields, although the density of
DeGrand-Toussaint monopoles~\cite{DeGrand:1980eq} is very small in
the LA gauge. The observation of
the dual Meissner effect in the LA gauge suggests that there exists a
gauge-independent definition of the monopole and,
consequently, of the monopole condensation.
There are some attempts to find a gauge-independent definition of
magnetic monopoles~\cite{fedor02,fedor05,maxim05}.
Based on the analogy of the SU(2) gluodynamics in the LA gauge with a nematic
crystal in Ref.~\cite{Chernodub:2005jh} an existence of various
topological defects was suggested.
But the definite answer about degrees of freedom which are relevant to the confinement
in the LA gauge is not yet obtained.

It has been shown that non-perturbative part of the  condensate
$\langle A_{\mu}^2(s) \rangle$ is explained completely in terms of monopoles in
compact QED in Landau gauge~\cite{fedor1}, where the monopole condensation
is known to be responsible for the confinement of charge~\cite{compactQED}.
The non-perturbative part of $\langle A_{\mu}^2(s) \rangle$ corresponds just to the
vacuum expectation value of a dual Higgs (monopole) field. For
gluodynamics the relevance of the $\langle A^2(s) \rangle$ condensate for confinement
is discussed in Refs.~\cite{fedor2,stodolsky}. Using the sum rule
technique the mass gap generation due to $d=2$ gluon condensate is
discussed in Ref.~\cite{Mass}.

In this note we fix the type of the vacuum also in the LA gauge. First we
measure the penetration length from the electric field flux as done in
the MA gauge. Then we fix the coherence length from correlations between
Wilson loops and the dimension 2 gluon operator $A^2(s)$, assuming that
a relation between the dimension 2 operator and an unknown gauge-independent monopole exists
in the LA gauge similarly to the MA gauge. We find both the penetration
length and the coherence length in the LA gauge are consistent with those in the
MA gauge. The type of the vacuum is found to be gauge-independent as it should be.
Note that the LA gauge corresponds to a gauge in which the functional $\sum_{s}A^2(s)$ is minimized.
Thus the operator $\sum_{s}A^2(s)$ could have a physical meaning in LA
gauge~\cite{stodolsky,kondo} if the Gribov-copy problem is solved.

\section{Consideration in the dual Ginzburg-Landau theory}
\subsection{General dual Ginzburg Landau picture}

The monopole density around the QCD string is described very well by the dual superconductor picture~\cite{bali98,ref:string:MA:gauge}.
The dual superconductor (or, the dual Ginzburg-Landau (DGL)) Lagrangian  corresponding to SU(2)
gluodynamics has the following form:
\beqn
\cL_{DGL} &=& \frac{1}{4} \, {(\partial_{[\mu,} B_{\nu]})}^2 + \frac{1}{2}
{|(\partial_\mu + i g\, B_\mu)\Phi|}^2 \nonumber\\
  && + \kappa {(|\Phi|^2 - \eta^2)}^2\,,
\label{eq:DLG}
\eeqn
where $B_\mu$ is the dual gauge field with the mass $m_B = g \eta$, and $\Phi$ is the monopole field with magnetic charge $g$
and with the mass $m_\Phi = \sqrt{8 \kappa} \eta$. In the confinement phase of SU(2) gluodynamics the monopoles
are condensed, $|<\Phi>| = \eta$. The coupling $\kappa$ defines the quartic interaction of the monopole field $\Phi$.
Below we discuss some general well-known properties of the Abrikosov-Nielsen-Olesen~\cite{ref:Abrikosov} vortex in this Abelian model.

There are two mass-scales in the discussed Abelian Higgs model: the coherence length $\xi$ and the penetration length $\lambda$, which
are inversely proportional to the masses of the monopole and the dual gauge boson, respectively:
\beqn
\xi = \frac{1}{m_\Phi}\,,\qquad \lambda = \frac{1}{m_B}\,.
\eeqn
The border between the type 1 and type 2 superconductors is defined as a region in the phase diagram space where the both length coincide,
$\xi = \lambda$.

We are interested in the behavior of the monopoles around the QCD string.
The classical equations of motion of the DGL model~\eq{eq:DLG} contain a solution corresponding to the QCD string with
a quark and an anti-quark at its ends. The infinitely separated quark and anti-quark correspond to an axially symmetric
solution of the string. We consider the static solution which is parallel to the third direction of the reference system,
\beqn
\Phi(\rho) &=& \eta f(\rho) \, e^{i \theta}\,,\quad \nonumber\\
B_i &=& \frac{\epsilon_{ij}}{g} \frac{x_j}{\rho^2} \, h(\rho)\,, \quad
B_3 = 0\,, \quad B_4 = 0\,,
\label{eq:anzatz}
\eeqn
where $f(\rho)$ and $g(\rho)$ are dimensionless functions of the transverse distance $\rho = \sqrt{x^2_1 + x^2_2}$ to the string, and
$\epsilon_{ij}$ is the standard fully anti-symmetric tensor,  $\epsilon_{ij}= - \epsilon_{ji}$ and $\epsilon_{12}=1$. The
azimuthal angle is\footnote{There is a mix of notations with the previous Section in which $\theta_{x,\mu}$ entered the
definition of the link matrix $U_{x,\mu}$. We believe this would not cause misunderstanding since this and the previous Sections
are independent.} $\theta \equiv \arg(x_1 + i x_2)$. Both functions $f$ and $h$ of Eq.~\eq{eq:anzatz} tend to zero
as $\rho\to 0$ and they approach the unity as $\rho \to \infty$.

The DGL classical equations of motion are:
\beqn
& & D_\mu^2(B) \Phi - 4 \kappa (|\Phi|^2 - \eta^2) \Phi =0\,,\label{eq:class:1}\\
& & (\partial^2_\alpha \delta_{\mu\nu} - \partial_\mu \partial_\nu) B_\nu = g\, k_\mu (\Phi,B)\label{eq:class:2}\,,
\eeqn
where $k_\mu$ is the monopole current given by the following expression:
\beqn
k_\mu = \Im m \bigl[\Phi^* D_\mu(B) \Phi \bigr] \equiv {|\Phi|}^2 (\partial_\mu  \arg \Phi + g B_\mu)\,,
\label{eq:current:analytical}
\eeqn
where $D_\mu(B) = \partial_\mu + i g\, B_\mu$ is the covariant derivative.

In terms of the functions $f$ and $h$ used in the ansatz~\eq{eq:anzatz}, the current~\eq{eq:current:analytical} is given by:
\beqn
k_i = - \eta^2 \frac{\epsilon_{ij} x_j}{\rho} \frac{f^2(\rho)}{\rho} \bigl[1 - h(\rho)\bigr]\,,\quad k_3 =0\,, \quad k_4 = 0\,.
\label{eq:current:anzatz}
\eeqn
To derive this equation one should use the relation $\partial\theta/\partial x_i = - \epsilon_{ij} x_j/\rho^2$.

In terms of the ansatz~\eq{eq:anzatz} the classical equations of motion~\eq{eq:class:1} are:
\beqn
&& f''(\rho) + \frac{f'(\rho)}{\rho} - \frac{f(\rho)}{\rho^2} [1 -h(\rho)]^2
\nonumber\\
&&\hspace{22mm}+ \frac{m^2_\Phi}{2} [1-f^2(\rho)]
f(\rho) = 0 \,, \label{eq:class:2-1}\\
&& h''(\rho) - \frac{h'(\rho)}{\rho} + m^2_B [1-h(\rho)] f^2(\rho) = 0 \,. \label{eq:class:1-1}
\eeqn
Expanding the profile functions at large $\rho$, $f(\rho) = 1 - \delta f(\rho)$ and $h(\rho) = 1 - \delta h(\rho)$, and
keeping only linear terms~\footnote{This is justified because at large $\rho$ we have $f(\rho)\to 1$ and $h(\rho)\to 1$.}
in Eq.~\eq{eq:class:1-1} and Eq.~\eq{eq:class:2-1}, we get the linearized classical equations of motion:
\beqn
&& \delta f''(\rho) + \frac{\delta f'(\rho)}{\rho} - m^2_\Phi \cdot \delta f(\rho) = 0 \,, \label{eq:class:2-2}\\
&& \delta h''(\rho) - \frac{\delta h'(\rho)}{\rho} - m^2_B \cdot \delta h(\rho)  = 0 \,, \label{eq:class:1-2}
\eeqn
which have the solutions:
\beqn
\delta f(\rho) = C_f K_0(m_B \rho)\,, \quad \delta h(\rho) = C_h \, m_\Phi \rho\, K_1(m_\Phi \rho)\,.
\eeqn
Here $K_n$ are the modified Bessel functions with the following asymptotic ($x\to \infty$) expansion:
\beqn
K_n(x) = \sqrt{\frac{\pi}{2 x}} e^{-x} \Bigl[1 + O\bigl(x^{-1}\bigr)\Bigr]\,.
\label{eq:expansion2}
\eeqn
For the string solution with a lowest non-trivial flux
the arbitrary coefficient $C_f$ is always equal to unity, $C_{f}=1$, while the
coefficient $C_h$ is equal to unity in the Bogomol'ny limit ({\it i.e.}, exactly on the border between
the type 1 and type 2 superconductivity), Ref.~\cite{ref:NO,ref:Bettencourt}. Since the numerical results suggest strongly
that the SU(2) gauge theory is close to the border, we set  $C_h=1$ in our qualitative discussion below.
Thus, the functions $h$ and $f$ at large values of $\rho$ behave as follows:
\beqn
f(\rho) &=& 1 - I_0 (m_\Phi \rho) + \dots\,,
\label{eq:expansion1_1}\\
\quad
h(\rho) &=& 1 - m_B \rho \, I_1 (m_B \rho) + \dots\,,
\label{eq:expansion1_2}
\eeqn

\begin{figure*}[!thb]
\begin{center}
\begin{tabular}{ccc}
\includegraphics[scale=0.55,clip=true]{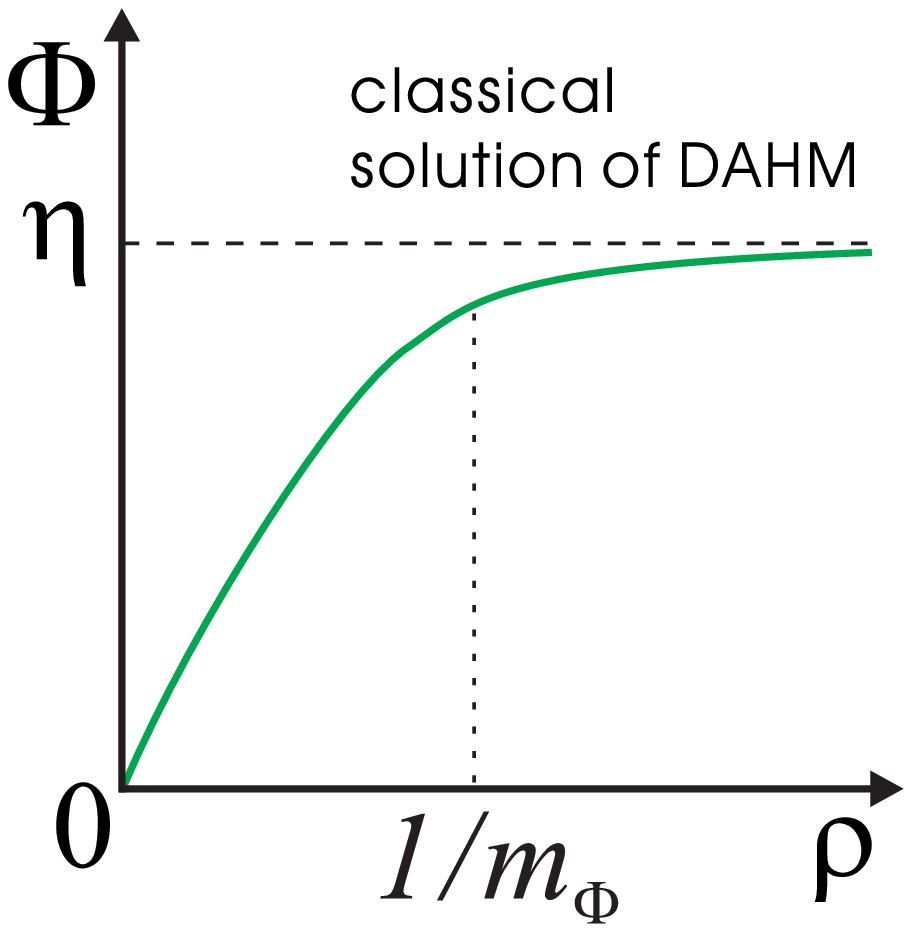}
\hspace{2mm} &
\includegraphics[scale=0.55,clip=true]{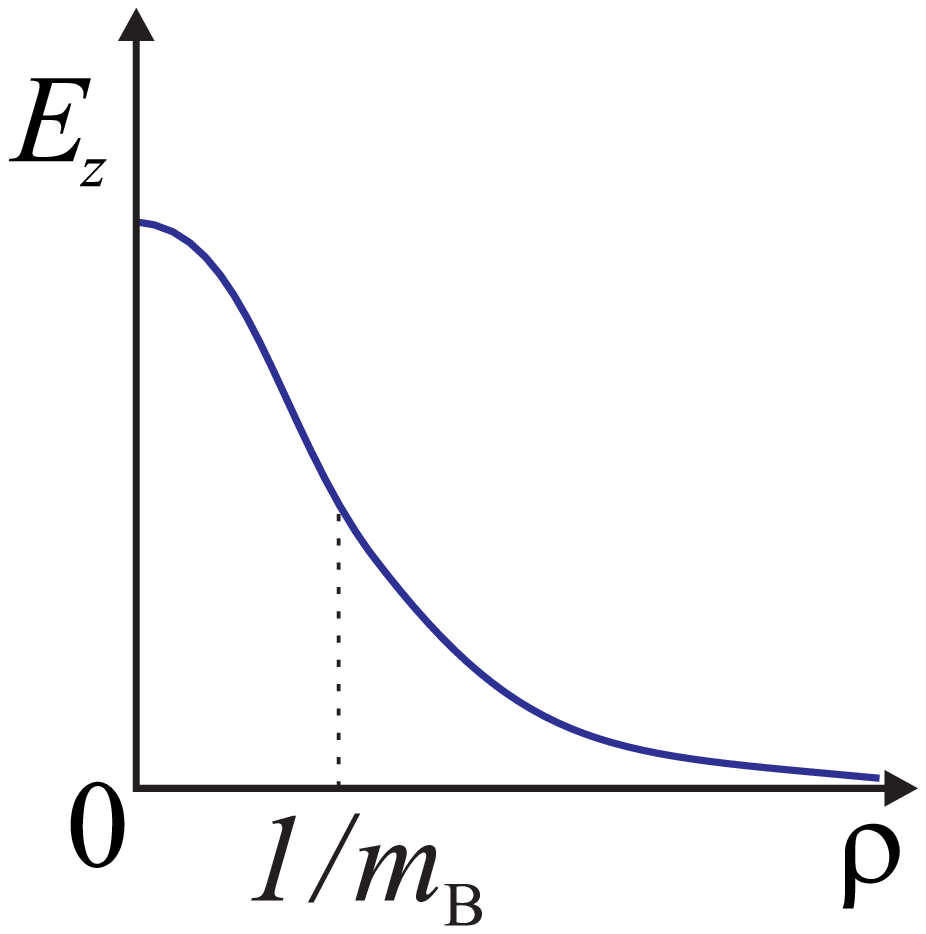}
\hspace{2mm} &
\includegraphics[scale=0.55,clip=true]{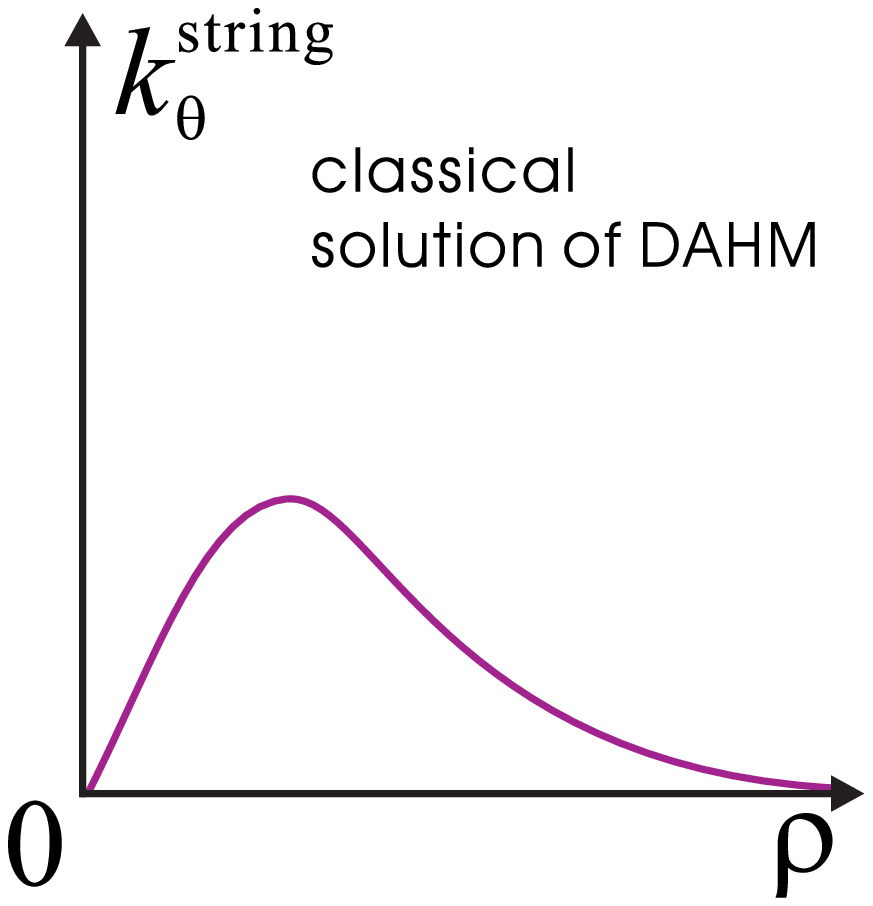} \\
(a) & (b) & (c)
\end{tabular}
\end{center}
\vspace{-4mm}
\caption{The qualitative behavior of (a) the monopole condensate $\Phi$, (b) the electric field $E_z$
and (c) the magnetic current component $k_\theta$ around the center of the string as functions of
the transverse distance $\rho$. The string is given by a classical solution of the DGL model.}
\label{fig:condensate:field}
\end{figure*}

The QCD string is well described by the solutions of the classical equations of motion of Lagrangian~\eq{eq:DLG}. The qualitative behavior
of the monopole condensate, the electric field and the angle component $k_\theta$ of the monopole current around the QCD string are
shown in Fig.~\ref{fig:condensate:field}(a), (b) and (c), respectively.

Summarizing, the value of the monopole current near the QCD string (obtained
from the classical equations of motion) is zero in the center of
the string and it is also zero far from the string.
The current has a maximum at a certain distance (which is numerically found to be about $0.2$~fm in the DLG corresponding to SU(2)
gluodynamics~\cite{bali98,ref:string:MA:gauge}).
The only non--zero component of the classical monopole current is the angle component
$k_\theta$, while other components (radial, temporal and $z$-component) are zero, $k_r=0$, $k_4=0$ and $k_3=0$.

\subsection{Monopole density around QCD string}

In the numerical calculations the distributions of the monopole current around the QCD string $\Sigma$ is measured with the help of the
correlation function
\beqn
k^{\mathrm{string}}_\mu = {\langle k_\mu(0) \rangle}_\Sigma \equiv \frac{\langle k_\mu(0) W_\cC\rangle }{\langle W_\cC\rangle}\,,
\qquad \partial \Sigma = \cC\,,
\label{eq:k:general}
\eeqn
where $\Sigma$ denotes the string world-sheet corresponding to the minimal surface spanned on the Wilson contour $\cC$.
The expectation value~\eq{eq:k:general} is non-zero contrary to Eq.\eq{eq:C:current} due to the broken Lorenz invariance because of
the presence of the string.

The monopole density is non-zero in the absence of the string. We
call this value of the density as "vacuum monopole density",
$|k^{\mathrm{vac}}|$. There are two contributions to this monopole
density coming from (i) the long (infrared) monopole loop which
forms the monopole condensate~\cite{ref:Hart,ref:Zakharov} and from (ii) the small monopole
loops which represent the perturbative (ultraviolet) fluctuations.

\begin{figure}[!thb]
\begin{center}
\includegraphics[scale=0.7,clip=true]{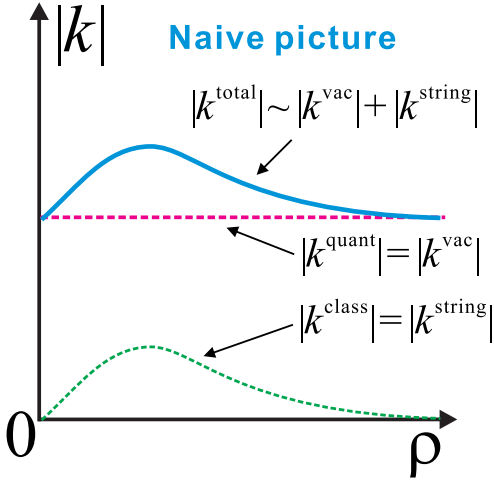}
\end{center}
\vspace{-4mm}
\caption{The naive qualitative behavior of the density of monopoles around the QCD string.}
\label{fig:condensate:field:naive}
\end{figure}

Very naively, the presence of the string should make the
monopole density bigger: the vacuum contribution gets an additional contribution coming from the
classical (solenoidal) current $k^{\mathrm{class}} \equiv k^{\mathrm{string}}$. The naive picture
is plotted in Fig.~\ref{fig:condensate:field:naive}.

Thus, naively, the density of the monopoles should increase at some distance from the string. Moreover,
naively one expects that at large transverse distance $\rho$ from the string the monopole density,
[according to Eqs.~(\ref{eq:current:anzatz},\ref{eq:expansion2},\ref{eq:expansion1_1},\ref{eq:expansion1_2})]
is controlled by the penetration length since
$|k^{\mathrm{string}}| = |k^{\mathrm{vac}}| + \mathrm{const} \, \exp\{ - m_B \rho\}$.

However, the described qualitative picture definitely contradicts~\footnote{Although the results of Ref.~\cite{ref:monopole:density}
were obtained in SU(3) gluodynamics, these results are applicable to our case since we do not expect
a qualitative difference between the behavior of the monopoles around the QCD string in the SU(3) and SU(2) cases.}
the numerical results obtained in Ref.~\cite{ref:monopole:density}.
In order to investigate the behavior of the monopole density near the QCD string we study analytically
the London limit in the next subsection.

\subsection{Monopole density in the vicinity of QCD string in the London limit}
\label{sec:london:limit}

The London limit is characterized by the infinitely deep potential, $m_\Phi \to \infty$.
The Lagrangian of the DGL model~\eq{eq:DLG} in this case is
\beqn
\cL^{\mathrm{London}}_{DGL} = \frac{1}{4} \, {(\partial_{[\mu,} B_{\nu]})}^2 + \frac{\eta^2}{2} {(\partial_\mu \varphi + g\, B_\mu)}^2\,.
\label{eq:DLG:London}
\eeqn
The QCD string $\Sigma$ manifests itself as a singularity in the phase of the Higgs field:
\beqn
\partial_{[\mu,}\partial_{\nu]} \varphi(x) & = & 2 \pi
{}^*\Sigma_{\mu\nu}(x)\,,\quad \nonumber \\
{}^*\Sigma_{\mu\nu} &=& \frac{1}{2} \epsilon_{\mu\nu\alpha\beta} \Sigma_{\alpha\beta}(x)\,,
\label{eq:string:varphi}
\eeqn
where the string $\Sigma$ is parameterized by the vector ${\bar x}_\mu(\tau_1,\tau_2)$ which
depends on the parameters $\tau_{1,2}$. The antisymmetric string current $\Sigma_{\mu\nu}$
is given by the following expression:
\beqn
\Sigma_{\mu\nu}(x) = \int \dd^2 \tau \,
\frac{\partial {\bar x}_{[\mu,}  \partial {\bar x}_{\nu]}}{\partial \tau_1 \partial \tau_2} \delta^{(4)}(x - {\bar x})\,.
\label{eq:string:parameter}
\eeqn
The partition function of the model~\eq{eq:DLG:London} can be rewritten as a sum over the closed strings~\footnote{Here and below we
drop all irrelevant constant pre-factors to the partition functions.}~\cite{ref:Akhmedov}:
\beqn
\cZ &=& \int^\pi_{-\pi} \dD \varphi \int_{-\infty}^\infty \dD B\,
\exp\Bigl\{ - \int \dd^4 \cL^{\mathrm{London}}_{DGL}(B,\varphi) \Bigr\}
\nonumber \\
 &=& \int_{\partial \Sigma = 0} \dD \Sigma \, \exp\bigl\{- S_{\mathrm{str}}(\Sigma)\bigr\}\,,
\label{eq:partition:1}
\eeqn
where $S_{str}$ is the string action:
\beqn
&&S_{\mathrm{str}}(\Sigma) = \nonumber\\
&&2 \pi^2 \eta^2 \int \dd^4 x  \int \dd^4 y \, \Sigma_{\mu\nu}(x) D_{m_B}(x - y) \Sigma_{\mu\nu}(y)\,,
\label{eq:string:action}
\eeqn
and $D_M$ is the propagator of the massive scalar particle, $(-\partial^2 +M^2) D_M(x) = \delta^{(4)}(x)$. The strings are closed:
$\partial_{\nu} \Sigma_{\mu\nu} = 0$. The derivation of the right hand side in Eq.~\eq{eq:partition:1} is easy to make
by fixing the unitary gauge, $\varphi=0$ and, consequently, making the shift $B_\mu \to B_\mu - (1/g) \partial_\mu \varphi$.
Then Eq.~\eq{eq:string:varphi} implies that under the shift
$\partial_{[\mu,} B_{\nu]} \to \partial_{[\mu,} B_{\nu]} - (2\pi/g) {}^*\Sigma_{\mu\nu}$. Finally,
having integrated over the Gaussian field $B_\mu$ we get the right hand side in Eq.~\eq{eq:partition:1}.

The sources of the electric flux ({\it i.e.}, the quarks) running along the trajectory $\cC$ are introduced with the help
of the Wilson loops written in terms of the original gauge fields $A_\mu$. The quantum average of the Wilson loop $W_\cC$
can be rewritten as a sum over the strings similarly to Eq.~\eq{eq:partition:1}:
\beqn
&&\langle W_\cC\rangle = \nonumber\\
&&\frac{1}{\cZ} \int_{\partial \Sigma = j^\cC}  \dD \Sigma \, \exp\bigl\{- S_{\mathrm{str}}(\Sigma)
- S_{\mathrm{current}}(j^\cC)\bigr\}\,,
\label{eq:partition:2}
\eeqn
where the strings are spanned on the current $j_\cC$: $\partial_{\nu} \Sigma_{\mu\nu} = j^\cC_\mu$. The action
for the currents is given by the short-ranged exchange of the dual gauge boson:
\beqn
&&S_{\mathrm{current}}(j_\cC) = \nonumber\\
&&\frac{e^2}{2} \int \dd^4 x  \int \dd^4 y \, j^\cC_\mu(x) D_{m_B}(x - y) j^\cC_\mu(y)\,,
\label{eq:electric:current:action}
\eeqn
where $e = 2\pi /g$ is the electric charge.

Below we evaluate the density of the monopole current in the
vicinity of the fixed QCD string. To this end we assume that
the leading contribution of the QCD string is naturally given by
the minimal surface configuration. Moreover, to avoid boundary (quark)
effects, we place the static quarks at the (spatial) infinities of
the axis $x_3$. Consequently, the quark term~\eq{eq:electric:current:action}
does not play any role in the forthcoming discussion.

Thus, we consider the infinite static string which is placed along the third direction. The
corresponding string current -- calculated from Eq.~\eq{eq:string:parameter} -- is given by the equation:
\beqn
\Sigma_{\mu\nu} = \Bigl(\delta_{\mu,3} \delta_{\mu,4} - \delta_{\nu,3} \delta_{\nu,4}\Bigr) \delta(x_1) \delta(x_2)\,.
\label{eq:static:string}
\eeqn
The monopole current~\eq{eq:current:analytical} in the London limit is
\beqn
k_\mu = \eta^2 (\partial_\mu \varphi + g B_\mu)\,.
\label{eq:current:London}
\eeqn

Let us consider the following generating functional:
\beqn
&& \cZ[\Sigma,C] =
\int_{-\infty}^\infty \dD B\, \nonumber\\
&& \times\, \exp\Bigl\{ - \int \dd^4 x\, \Bigl[\cL^{\mathrm{London}}_{DGL}(B,\varphi_\Sigma)
- i k_\mu C_\mu \Bigr]\Bigr\}\,,
\label{eq:partition:3}
\eeqn
where the singular phase $\varphi_\Sigma$ corresponds to the string position $\Sigma$ fixed via Eq.~\eq{eq:string:varphi}.
The monopole current in the presence of the string is given by the variational derivative:
\beqn
{\langle k_\mu(x) \rangle}_\Sigma = \frac{1}{\cZ[\Sigma,0]}\, \frac{\delta}{i \delta C_\mu(x)} \cZ[\Sigma,C] {\Bigl|}_{C=0}\,.
\label{eq:current:London:vev}
\eeqn
Analogously, the (squared) monopole density is
\beqn
{\langle k^2_\mu(x) \rangle}_\Sigma = \frac{1}{\cZ[\Sigma,0]}\,{\Bigr(\frac{\delta}{i \delta C_\mu(x)}\Bigr)}^2 \cZ[\Sigma,C] {\Bigl|}_{C=0}\,.
\label{eq:current2:London:vev}
\eeqn

Proceeding similarly to the derivation along Eqs.(\ref{eq:partition:1},\ref{eq:partition:2})
we get the following expression for the generating functional:
\beqn
&&\cZ[\Sigma,C] \nonumber\\
& = & \exp\Bigl\{ - \int \dd^4 x  \int \dd^4 y \, \Bigl[ \frac{g^2 \eta^4}{2} C_\mu(x) D^{m_B}_{\mu\nu}(x - y) C_\nu(y)
\nonumber\\
&- &  2 \pi i \eta^2 C_\mu(x) D^{m_B}_{\mu\nu}(x - y) \partial_\alpha  {}^* \Sigma_{\alpha\nu}(y)\Bigr]
- S_{\mathrm{str}}(\Sigma)\Bigr\}\,,
\eeqn
where $D^m_{\mu\nu}(x)$ is the propagator of the massive vector boson $B_\mu$, and the string action is given in Eq.~\eq{eq:string:action}.

An evaluation of the vacuum expectation value of the monopole density~\eq{eq:current:London:vev} gives:
\beqn
k^{\mathrm{string}}_\mu &\equiv& \langle k_\mu \rangle_\Sigma
\nonumber\\
&=& - 2 \pi \eta^2 \int \dd^4 y
\, D^{m_B}_{\mu\nu}(x - y) \partial_\alpha  {}^* \Sigma_{\alpha\nu}(y)\,.
\eeqn
In particular, in the case of
the static string~\eq{eq:static:string} we get the classical London solution:
\beqn
k^{\mathrm{string}}_i &=& - 2 \pi \eta^2 \epsilon_{ij} \frac{x_j}{\rho} \frac{\partial}{\partial \rho} D^{(2D)}_{m_B}(\rho)\,,
\quad i,j=1,2\,,\quad \nonumber\\
k^{\mathrm{string}}_3 &=&0\,,\quad k^{\mathrm{string}}_4 =0\,,
\label{eq:London:current}
\eeqn
where
\beqn
D^{(2D)}_{m_B} = \frac{1}{2\pi} K_0(m_B \rho)\,.
\label{eq:D2D}
\eeqn
is the propagator for a scalar massive particle in two space-time dimensions. Using Eqs.(\ref{eq:London:current},\ref{eq:D2D})
we get the explicit form of the only non-zero component of the solenoidal current:
\beqn
k^{\mathrm{string}}_\theta = \eta^2 m_B K_1(m_B \rho)\,.
\label{eq:London:current:general}
\eeqn
The monopoles form a solenoidal current which circulates around the string in transverse directions.

The squared monopole density is:
\beqn
\langle k^2_\mu \rangle_\Sigma = \langle k_\mu \rangle_\Sigma^2 + \langle k^2_\mu \rangle_0\,,
\eeqn
where
\beqn
(k^{\mathrm{quant}}_\mu )^2 &\equiv& \langle k^2_\mu \rangle_0
= g^2 \eta^4 D^{m_B}_{\mathrm{reg}}(0) \nonumber\\
&=& \frac{g^2 \eta^4 \Lambda^2}{16 \pi^2} + O\bigl(\log(\Lambda/m_B)\bigr)
\label{eq:density:quantum:correction}
\eeqn
is the quantum vacuum correction.
We have regularized the divergent expression of the vacuum correction by the momentum cutoff $\Lambda$.
The correction is quadratically divergent.

The total (squared) density of the monopole current is given by
\beqn
\langle k^2_\mu \rangle_\Sigma & = & \eta^4 M^2_B K^2_1(m_B \rho) \nonumber \\
& & + \frac{g^2 \eta^4 \Lambda^2}{16 \pi^2} +
O\bigl(\log(\Lambda/m_B)\bigr)\,,
\label{eq:current:main}
\eeqn
where the solenoidal current $k^{\mathrm{string}}$ in the London limit is given by Eq.\eq{eq:London:current}.
This expression is the {\it exact} in the London limit (up to logarithmically divergent but
distance-independent corrections).

One may easily see from Eq.~\eq{eq:current:main} that the naive expectation of the density behavior --
shown in Fig.~\ref{fig:condensate:field:naive} -- is, in fact, correct in the London limit. Then
the total density (in which the coherence length is zero) must have a maximum at the distance of the order of
the penetration length, $1/m_B$. However, the naive picture depicted in Fig.~\ref{fig:condensate:field:naive}
is not valid the case of the finite coherence length considered below.

\subsection{Monopole density  in the vicinity of QCD string for finite coherence length}
\label{sec:real:case}

Here we show, that in the real case os a finite coherence length, the
naive picture, described in the previous Subsection, is no more correct.
Indeed, in this case the value of the monopole condensate is varying in
the vicinity of the string, and the (qualitative, at least) generalization of
Eq.~\eq{eq:current:main} should read as follows:
\beqn
\langle k^2_\mu \rangle_\Sigma &\equiv& (k^{\mathrm{string}}_\mu)^2 +
(k^{\mathrm{quant}}_\mu )^2  \nonumber\\
&=&
(k^{\mathrm{string}}_\mu)^2 + \frac{g^2 |\Phi(\rho)|^4 \Lambda^2}{16 \pi^2} + \dots\,,
\label{eq:current:main:general:real}
\eeqn
where we have taken into account the behavior of the monopole condensate by the simple replacement~\footnote{Clearly, in a
full and accurate treatment of the problem one should also consider the renormalization of the quantum corrections due to
the varying condensate. Our considerations in this Section are of a qualitative nature therefore we skip the discussion
of the renormalization.} $\eta \to |\Phi(\rho)|$ in
\beqn
(k^{\mathrm{quant}}_\mu )^2 \equiv \langle k^2_\mu \rangle_0 = \frac{g^2 |\Phi(\rho)|^4 \Lambda^2}{16 \pi^2} + \dots\,.
\eeqn
Note that the quantum correction to the squared monopole current in the vicinity of the string (with $\rho \sim \xi$) is
not equal to the vacuum expectation value measured far outside the string ($\rho \gg \xi$)!

\begin{figure}[!thb]
\begin{center}
\includegraphics[scale=0.75,clip=true]{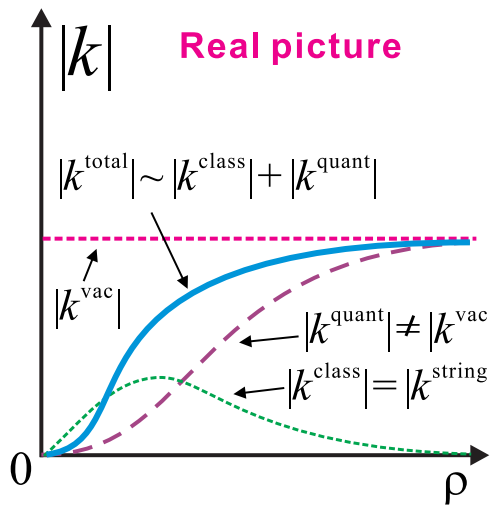}
\end{center}
\vspace{-4mm}
\caption{The real qualitative behavior of the density of monopoles around the QCD string.}
\label{fig:condensate:field:real}
\end{figure}

The quantum correction is much stronger than the classical one, therefore the leading behavior of the total density is
controlled by the quantum corrections. The behavior of the monopole density in the vicinity of the string is shown in
Fig.~\ref{fig:condensate:field:real} by the solid line.
The various contributions to the total density are also shown in this Figure (the dashed lines). The theoretical
expectation -- shown in Fig.~\ref{fig:condensate:field:real} -- is in agreement with the numerical
result of Ref.~\cite{ref:monopole:density}.

Thus, we expect that the quantum corrections play an essential role in the case of a finite coherence length. Moreover,
the leading behavior of the monopole density at large distances is controlled by the coherence length $\xi$ (and not by the penetration
length $\lambda$). This fact can be seen from Eqs.~(\ref{eq:expansion2},\ref{eq:expansion1_1},\ref{eq:expansion1_2}) in the limit
$\rho \gg \xi$:
\beqn
\langle |k| \rangle_\Sigma(\rho) &\sim& {\bigl(\langle k^2_\mu
\rangle_\Sigma\bigr)}^{1/2}(\rho) \nonumber\\
&=&
\frac{g^2 \Lambda^2}{16 \pi^2} \Biggl[1 - 4 \sqrt{\frac{\pi \xi}{2 \rho}} \, e^{ - \rho/\xi}\Biggr] + \dots\,.
\label{eq:pre-final}
\eeqn

As it is discussed in Section~\ref{sec:A:monopole:correlations},
the monopole density should locally be correlated with the
condensate $A^+ A^-$ (one can naturally expect a correlation with a short length scale
of the order of 0.05fm -- this topic will be discussed in the next Section).
Therefore, the correlation of the monopole density with the QCD string~\eq{eq:pre-final}
indicates that the $A^+ A^-$  condensate is also correlated with the QCD
string. The correlation lengths for the "monopole
density"--"string" correlations and for the "$A^+ A^-$
condensate"--"string" correlations should be the same and equal to the coherence length of the
dual superconductor:
\beqn
\frac{\langle A^+_\mu(\rho) A^-_\mu(\rho)\rangle_\Sigma}{\langle A^+_\mu(\rho) A^-_\mu(\rho)\rangle_0}
= 1 - {\mathrm{const}}\cdot e^{ - \rho/\xi} + \dots\,.
\label{eq:A2:general}
\eeqn
with $\rho \gg \xi$. This is the main result of this Section.

\section{$A^+ A^-$ condensate and Abelian monopole in MA gauge}
\label{sec:A:monopole:correlations}

The MA gauge is defined by the maximization of the functional
\beqn
R[U] = \sum_l R_l[U]\,,\qquad R_l[U] = \frac{1}{2} {\mathrm{Tr}}\Bigl[U_l \sigma_3 U_l^\dagger \sigma_3 \Bigr]\,,
\eeqn
with respect to the gauge transformations,
\beqn
\max_{\Omega \in SU(2)} R[U^\Omega]\,, \qquad U^\Omega_{x,\mu} = \Omega^\dagger_x U_{x,\mu} \Omega_{x+\hat \mu}\,.
\eeqn
Using the standard parameterization of the $SU(2)$ link matrices,
\beqn
U_l = \left(
\begin{array}{ll}
\cos \phi_l \, e^{i \theta_l} & \sin \phi_l \, e^{i \chi_l} \\
- \sin \phi_l \, e^{- i \chi_l} & \cos \phi_l \, e^{- i \theta_l}\\
\end{array}
\right)\,, \label{projection}
\eeqn
we obtain $R_l[U] = \cos 2 \phi_l$. The maximization makes $\phi$ as close to zero as possible.

The off-diagonal fields, $U^\pm_l = \pm \sin \phi_l \, e^{\pm i \chi_l}$ correspond to continuum fields $\pm A^\pm_\mu(x)$ and
continuum quantity $A^+_\mu(x)A^-_\mu(x)$ corresponds to the lattice
quantity $\sin^2 \phi_l \equiv (1 - R_l[U])/2$. Thus we are able to
make an identification (no summation over $\mu$ is assumed):
\beqn
A^+_\mu(x)A^-_\mu(x) = \frac{1}{2} \bigl(1 - R_{x,\mu}[U]\bigr)\,,
\label{eq:A:R}
\eeqn
where the equality is exact in the naive continuum limit.

The first measurements of the local correlation between monopoles and the quantity $R_l$ were done in
Ref.~\cite{ref:ambiguity}, where the quantity $R_c$ was used:
\beqn
R_c[U] = \sum_{l \in \partial c} R_l[U]\,.
\label{eqR_c}
\eeqn
The summation is going over all links belonging to the cube $c$. The distribution of the quantity $R_c$
at the cubes occupied by monopoles and the cubes, not occupied by monopoles, is observed.
It is shown that at the monopole position the quantity $R_c$ is generally smaller compared to the
same quantity in the empty space. Therefore, the monopoles suppress the quantity $R_c$, and, according
to Eq.~\eq{eq:A:R}, the $A^+A^-$ condensate is enhanced on monopoles.

One can suggest, that the correlation of the $A^+A^-$ condensate with the monopole is short ranged.
Indeed, the correlation of the monopole with the SU(2) action in the MA gauge is short ranged, with the
characteristic correlation length~\cite{ref:anatomy} $\zeta_{\mathrm{Action}} \approx 0.05$fm. Since the SU(2) action involves
the off-diagonal components, it seems natural to suggest that the correlation length $\zeta_{\mathrm{cond}}$ of
the off-diagonal components of the gluon field $A^\pm$ (or, of the $A^+A^-$ condensate) with the monopoles is not much
higher than the $\zeta_{\mathrm{Action}}$. Thus, one can expect that $\zeta_{\mathrm{cond}}
\approx \zeta_{\mathrm{Action}} \approx 0.05$fm.

Thus the $A^+A^-$--monopole density correlation function,
\beqn
C(r) = \frac{\langle |k_\nu(0)| A^+_\mu(r) A^-_\mu(r)\rangle }{\langle |k_\nu(0)| \rangle \langle A^+_\mu(0) A^-_\mu(0)\rangle}-1\,,
\label{eq:C:general}
\eeqn
at large $r$ is an exponentially decaying function with characteristic length scale $\zeta$.

In general there are two types of correlations: along the monopole
current $k_\mu$ and perpendicular to the monopole current. In
Eq.~\eq{eq:C:general} we assume that the distance $\vec r$ is
perpendicular to the direction of the monopole current, $R_\mu
k_\mu =0$ ({\it i.e.}, the correlations are studied in the transverse to the
monopole current directions). Obviously, due to the scalar nature of the $A^+A^-$
operator the correlation of this quantity with the monopole current is zero:
\beqn
\langle k_\nu(0) A^+_\mu(r) A^-_\mu(r)\rangle \equiv 0\,.
\label{eq:C:current}
\eeqn

\section{Numerical results}
\subsection{Method}

We use an improved gluonic action found by Iwasaki~\cite{iwasaki}
which was already implemented in Ref.~\cite{suzuki05}:
$$
S = \beta \left\{C_0\! \sum {\mathrm{Tr}} \mbox{(plaquette)} + C_1\! \sum {\mathrm{Tr}} \mbox{(rectangular)} \right\}.
$$
The mixing parameters are fixed as $C_0 + 8 C_1 = 1$ and $C_1=-0.331$.
We adopt the coupling constant $\beta=1.2$ which corresponds to
the lattice distance $a(\beta=1.2) = 0.0792(2)$fm. The lattice size is $32^4$
and we use around  5000 thermalized configurations for measurements.
To get a good signal-to-noise ratio, the APE smearing technique~\cite{APE}
is used when evaluating Wilson loops $W(R, T)=W^0+iW^a\sigma^a$.
The thermalized vacuum configurations are gauge-transformed in the MA($+$ U1LA)
gauge and in the LA gauge. In the LA gauge the functional
$\sum_{s,\mu} Tr [ U_{\mu}(s) + U_{\mu}^{\dagger}(s)]$ is maximized with respect to
all gauge transformations.

\subsection{The MA gauge case}
\begin{figure}[htb]
\includegraphics[height=6.5cm, width=8.5cm]{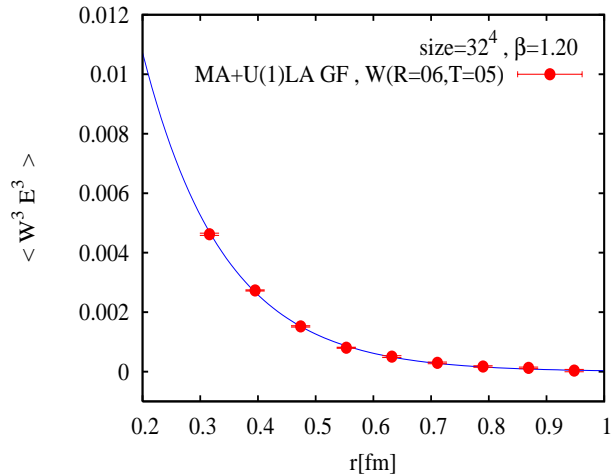}
\caption{\label{fig_1_ma} The non-Abelian $\vec{E}$
 electric field profile in the MA $+$ U1LA gauge obtained with the use of the $R\times T= 6\times 5$
 Wilson loop. The solid line denotes the best exponential fit by
 the function~\eq{eq:fit} with the best fit parameters given in Table~\ref{tbl:best:fits}.}
\end{figure}

Non-Abelian electric fields  are defined from $1\times 1$ plaquette
$U_{\mu\nu}(s)=U^0_{\mu\nu}(s)+iU^a_{\mu\nu}(s)\sigma^a$ as done in Ref.~\cite{bali-94}:
\begin{eqnarray}
E^a_k(s)&\equiv& \frac{1}{2}[U^a_{4k}(s-\hat{k})+U^a_{4k}(s)]
\label{eq:electric:field}
\end{eqnarray}
The static quarks are represented by the Wilson loop
$W(R,T)$. The measurements of the electric field are mainly done on the
perpendicular plane at the midpoint between the quark pair. A typical
example is shown in Fig.~\ref{fig_1_ma}. Note that electric fields
perpendicular to the $Q\bar{Q}$ axis are found to be negligible.

\begin{table*}[!htb]
\begin{tabular}{|c|c|c|c|c|c|c|}
\hline
Quantity                          &  gauge      & Fig.            & $\zeta$ [fm]      & $c_1$       & $c_2$           & $\chi^2/d.o.f.$ \\
\hline
$\langle W^3 E^3 \rangle           $ &  MA + U1LA  & \ref{fig_1_ma}  & $0.140(3)$   &  $0.044(2)$ &  $0$            & $3.06$    \\
$\langle |k_\mu| (A^+A^-)_u \rangle    $ &  MA         & \ref{k-AA}      & $0.0606(9)$  &  $1.08(3)$  &  $-0.0103(2)$   & $0.003$   \\
$\langle W  k^2 \rangle            $ &  MA         & \ref{w-mono}    & $0.10(1)$    &  $0.016(7)$ &  $0.014300(8)$  & $1.35$    \\
$\langle W (A^+A^-)_u \rangle      $ &  MA         & \ref{w-AA}      & $0.094(8)$   &  $0.04(1)$  &  $0.40967(2)$   & $0.006$   \\
$\langle W (A^+A^-)_\theta \rangle $ &  MA + U1LA  & \ref{w-AA_theta}& $0.106(9)$   &  $0.12(4)$  &  $0.49068(4)$   & $0.01$    \\
$\langle W^3 E^3 \rangle           $ &  LA         & \ref{fig_1}     & $0.139(1)$   &  $0.038(1)$ &  $0$            & $1.88$    \\
$\langle W A^2_\theta \rangle      $ &  LA         & \ref{fig_4}     & $0.118(4)$   &  $0.09(1)$  &  $0.74023(4)$   & $0.003$   \\
\hline
\end{tabular}
\caption{\label{tbl:best:fits} The best fit parameters corresponding to the fits of various quantities by the function~\eq{eq:fit}. We indicate
the gauge where the quantity is calculated and the figure number where the quantity is plotted. We set $c_2=0$ when the best fit value
of $c_2$ is consistent with zero.}
\end{table*}

The correlation length is determined by an exponential fit of the electric
field~\ref{eq:electric:field} for large $r$ regions, where $r$ is a distance perpendicular to the $Q\bar{Q}$ axis.
Below we observe that the electric field as well as other field distributions around the string
can be fitted well by a simple exponential function
\beqn
f(r)=c_1 \exp(-r/\zeta)+c_2\,,
\label{eq:fit}
\eeqn
where $\zeta$ and $c_{1,2}$ are the fit parameters. The corresponding best fit parameters are presented
in Table~\ref{tbl:best:fits}. The best fitting curve for the distribution of the
electric field is plotted in Fig.~\ref{fig_1_ma} as solid line. From this fit
we fix the penetration length\footnote{We have fixed both lengths using a simple exponential function~\eq{eq:fit}
expected to work well in the long-range region.
For short-range regions, the function is not suitable, so that we have omitted the first three or four points.
Changing the fitting range, we found the fitted length tends to be smaller and then to be rather stabilized for some
sets of range and then again becomes smaller. We choose the value at the stabilized range
and consider the change of the fitted values as a systematic error. Hence all error bars in the figures here
with respect to lengths include such systematic errors in addition to the statistical errors.}
$\lambda$.

\begin{figure}[htb]
\includegraphics[height=6.5cm, width=8.5cm]{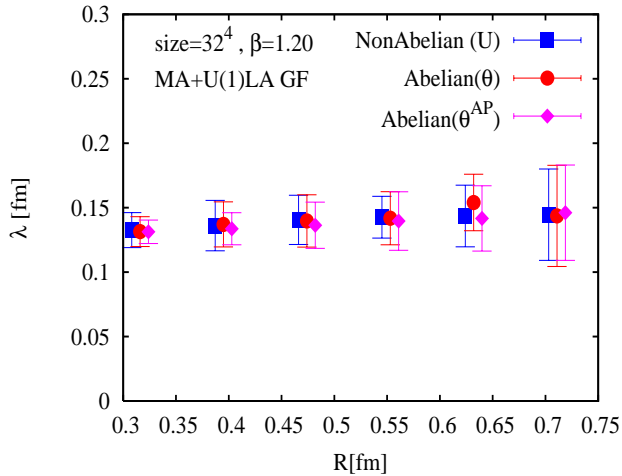}
\caption{\label{fig_3} The penetration lengths for non-Abelian $\vec{E}$ and Abelian $\vec{E}_A$($\theta$ and
$\theta^{AP}$) electric field  profiles in the MA $+$ U1LA gauge and their $R$-dependence.
Here $\theta$ is defined by $U_\mu(s)=\exp(i\theta_\mu^a(s)\sigma^a)$ and
$\theta^{AP}$ is given in Eq.~(\ref{projection}).}
\end{figure}

We show the results for the penetration length in the MA $+$ U1LA gauge
in Fig.~\ref{fig_3} for various sizes of Wilson loop in space directions $R$. Here we see the penetration
lengths for both Abelian $\vec{E}_A$ and non-Abelian $\vec{E}$ electric fields are compatible with each other.
This is expected, since in MA gauge off-diagonal gluon components are forced to become as small as possible.

\begin{figure}[htb]
\includegraphics[height=6.5cm, width=8.5cm]{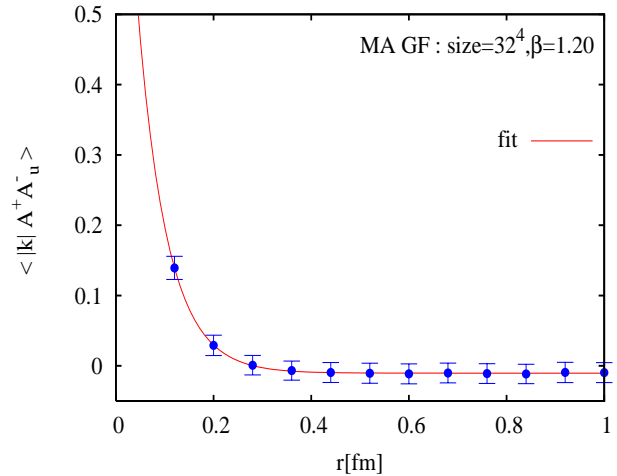}
\caption{\label{k-AA} The correlation between the monopole density
$|k_\mu|$ and the operator $A^+A^-$ in the MA gauge.
 The solid line denotes the best exponential fit by
 the function~\eq{eq:fit} with the best fit parameters given in Table~\ref{tbl:best:fits}.}
\end{figure}

Next we study the correlation between the monopole density $\left|k_\mu(s)\right|$
and the operator  $A^+A^-(s)$ given by Eq.~(\ref{eq:A:R}).
The correlation data is plotted in Fig.~\ref{k-AA}. It is completely consistent with the theoretical
expectation discussed in Section~\ref{sec:A:monopole:correlations}. In particular, the scale of correlations
between $\left|k_\mu(s)\right|$ and $A^+A^-(s)$ is about $0.06$~fm according to Table~\ref{tbl:best:fits}. This value
is pretty close to the scale $\zeta_{\mathrm{Action}} \approx 0.05$fm of the monopole-action correlations.

\begin{figure}[htb]
\includegraphics[height=6.5cm, width=7.5cm]{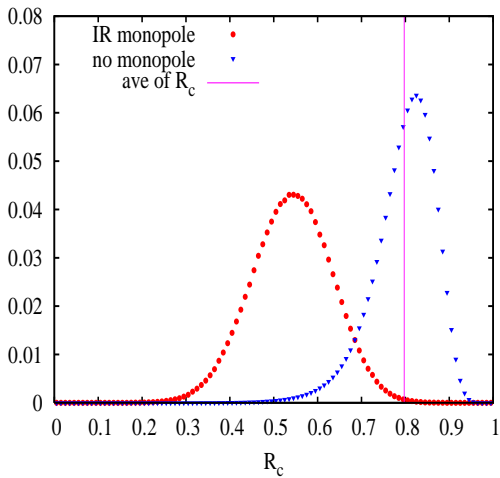}
\caption{\label{hist-new} The histogram of the distribution of the
quantity $R_c$ in Eq.(\ref{eqR_c}) at the locations of infrared monopoles and outside the monopole locations.
The vertical line indicates the vacuum average $\langle R_c\rangle$.}
\end{figure}

The histograms of the quantity $R_c$, Eq.(\ref{eqR_c}), are plotted in Fig.~\ref{hist-new}.
We discriminate between the histograms obtained at the cubes unoccupied by monopoles and obtained at
the cubes occupied by long infrared monopoles. From this figure, we can clearly see the enhancement of
the $A^+A^-$ condensate on the Abelian monopoles.

%fig7

%fig8

\begin{figure}[htb]
\includegraphics[height=6.5cm, width=8.5cm]{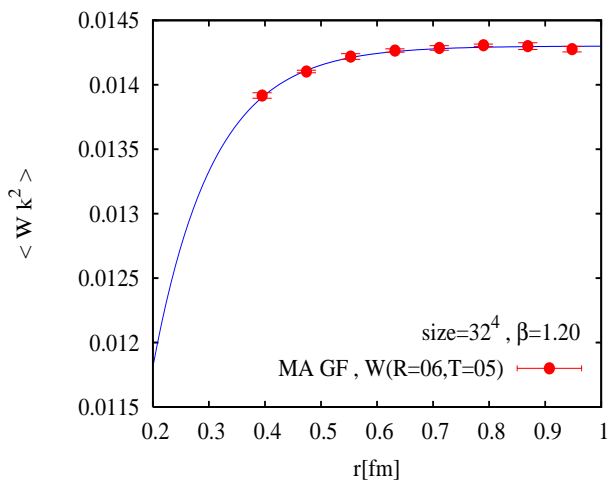}
\caption{\label{w-mono} The correlation between the $R\times T= 6\times 5$ Wilson loop and the
square monopole density in the MA gauge. The solid line denotes the best exponential fit by
 the function~\eq{eq:fit} with the best fit parameters given in Table~\ref{tbl:best:fits}.}
\end{figure}

\begin{figure}[htb]
\includegraphics[height=6.5cm, width=8.5cm]{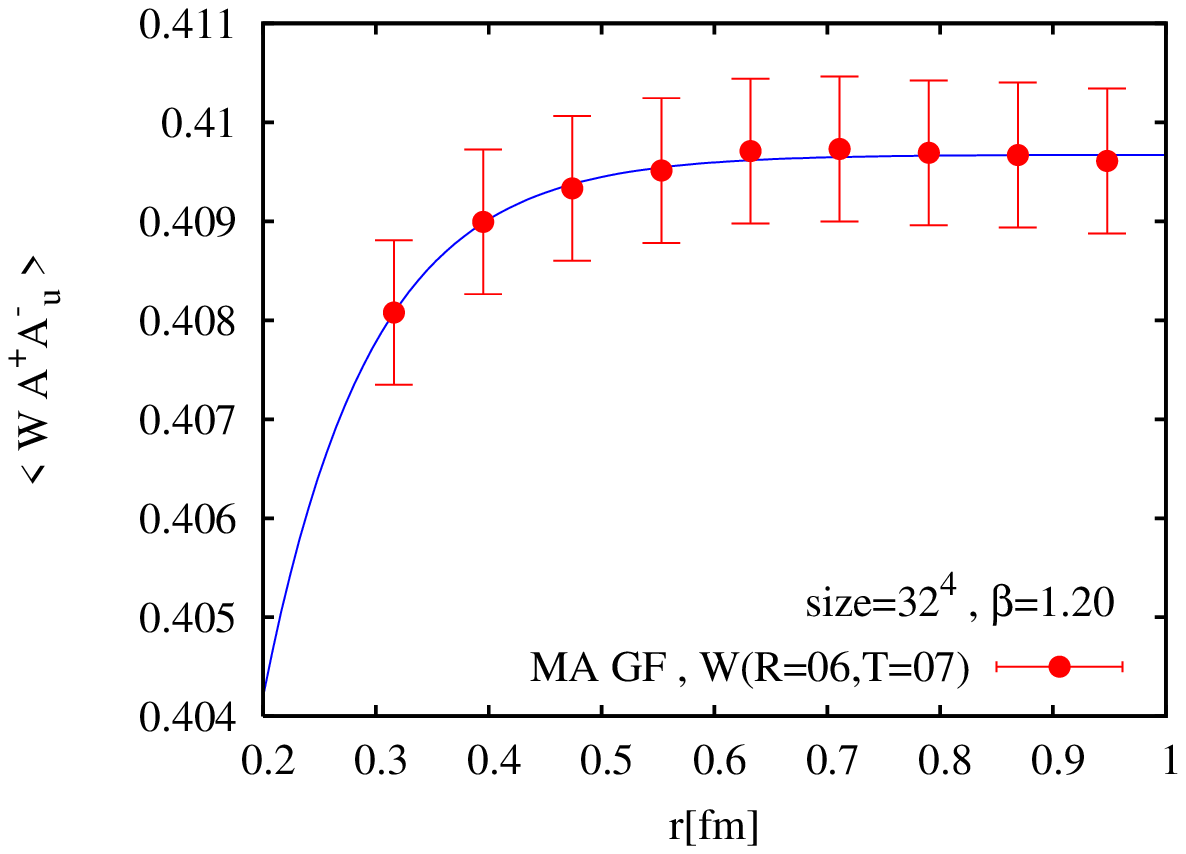}% Here is how to import EPS art
\caption{\label{w-AA} The same as in Fig.~\ref{w-mono} but for the correlation between the $R\times T= 6\times 7$ Wilson loop and
 the $A^+A^-_u$ condensate~\eq{eq:A:R}.}
\end{figure}

Let us next  derive the coherence length in the MA gauge.
The correlations between the Wilson loop and the monopole density, and
between the Wilson loop and the quantity $A^+A^-(s)$ are plotted in
Fig.~\ref{w-mono} and in Fig.~\ref{w-AA}, respectively.
\begin{figure}[htb]
\includegraphics[height=6.5cm, width=8.5cm]{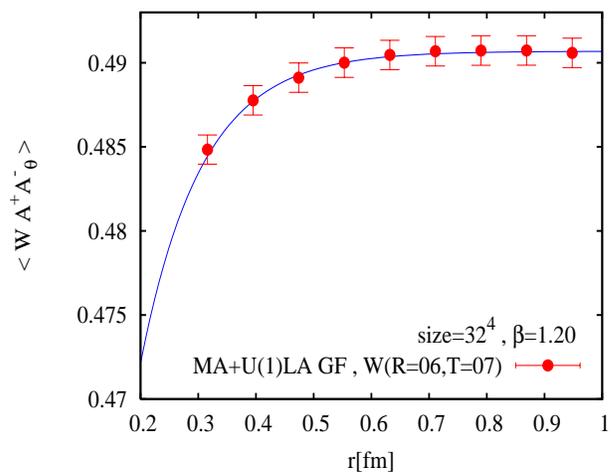}% Here is how to import EPS art
\caption{\label{w-AA_theta} The same as in Fig.~\ref{w-mono} but for the correlation between the $R\times T= 6\times 7$ Wilson loop and
 the $A^+A^-_\theta$ condensate~\eq{eq:AA_theta} in the MA $+$ U1LA gauge.}
\end{figure}
\begin{figure}[htb]
\includegraphics[height=6.5cm, width=8.5cm]{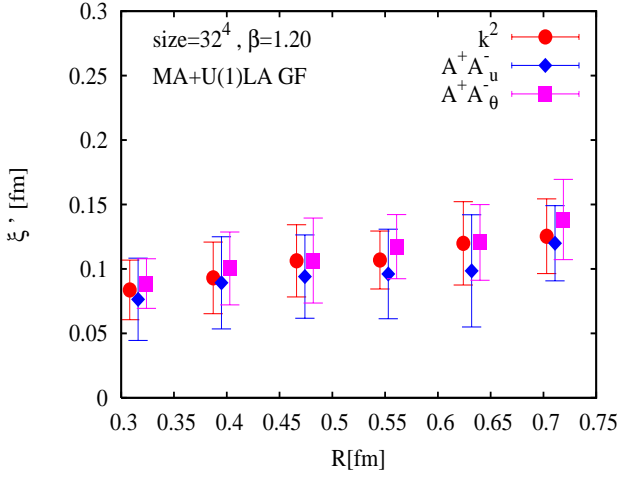}
\caption{\label{w-AAk2_coherence} The coherence lengths $\xi'$ for $A^+A^-_u$ and
$A^+A^-_\theta$ and $k^2$ in the MA $+$ U1LA gauge.}
\end{figure}
\begin{figure}[htb]
\includegraphics[height=6.5cm, width=8.5cm]{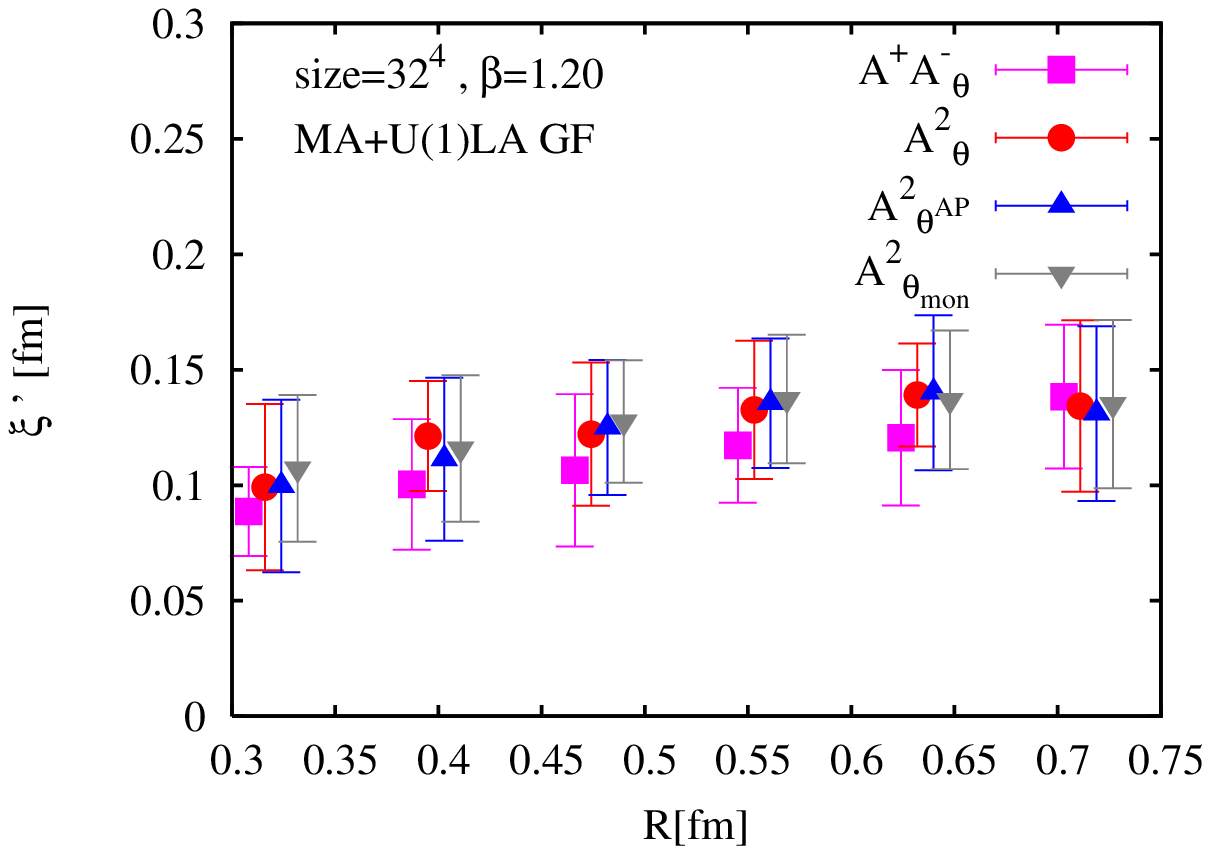}
\caption{\label{fig_6} The coherence lengths of dimension 2 gluon operator in the MA $+$ U1LA gauge for
various $R$.}
\end{figure}
\begin{figure}[htb]
\includegraphics[height=6.5cm, width=8.5cm]{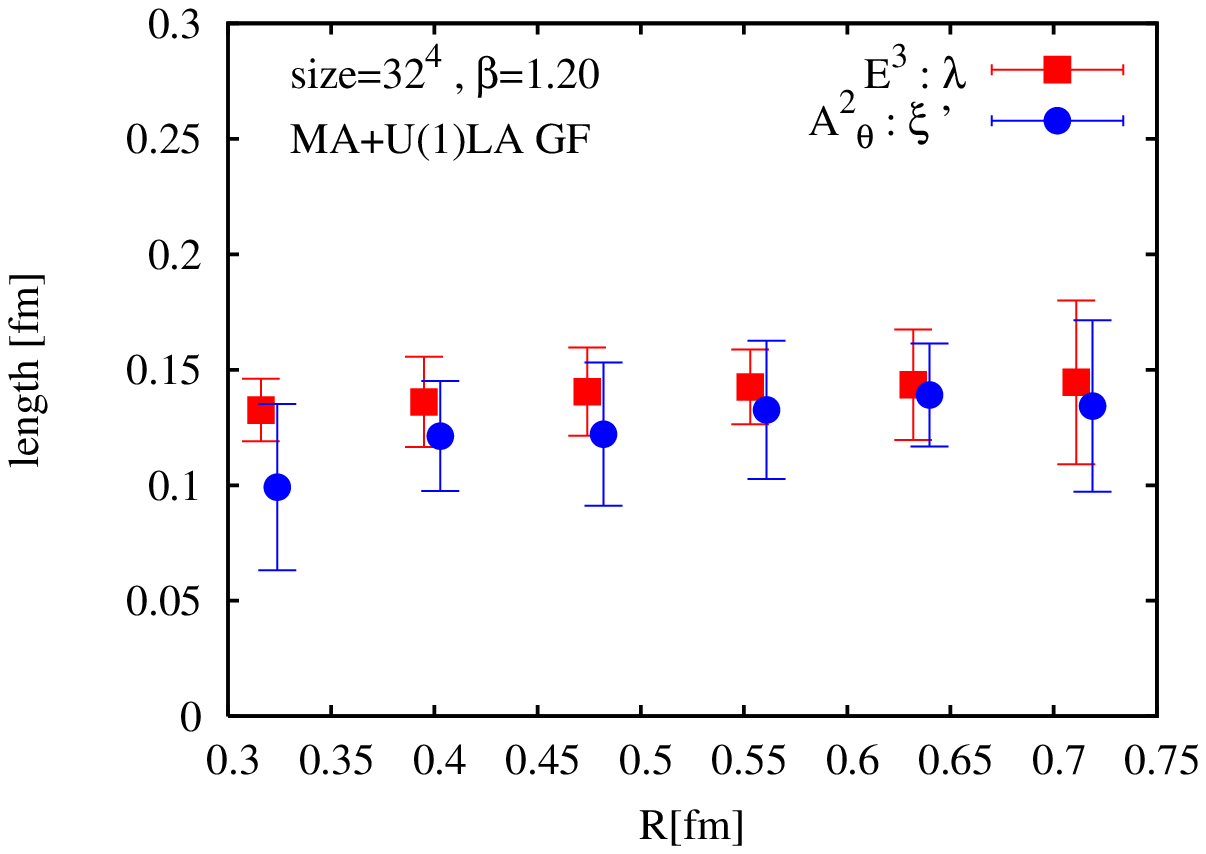}% Here is how to import EPS art
\caption{\label{fig_cp_ma}
The coherence lengths of dimension 2 gluon operator and the
penetration length in MA $+$ U1LA gauge for various $R$.}
\end{figure}
\begin{figure}[htb]
\includegraphics[height=6.5cm, width=8.5cm]{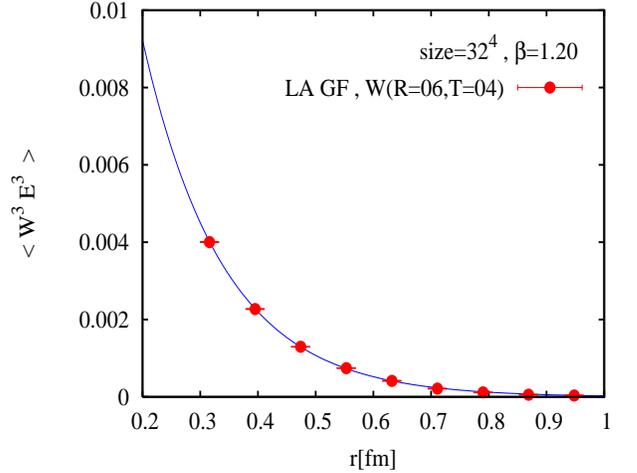}% Here is how to import EPS art
\caption{\label{fig_1}
The non-Abelian $\vec{E}$ electric field  profile in the Landau gauge obtained with the help of the Wilson loop $W(R\times T= 6\times 4)$.
The solid line denotes the best exponential fit by the function~\eq{eq:fit} with the best fit parameters given in Table~\ref{tbl:best:fits}.}
\end{figure}
\begin{figure}[htb]
\includegraphics[height=6.5cm, width=8.5cm]{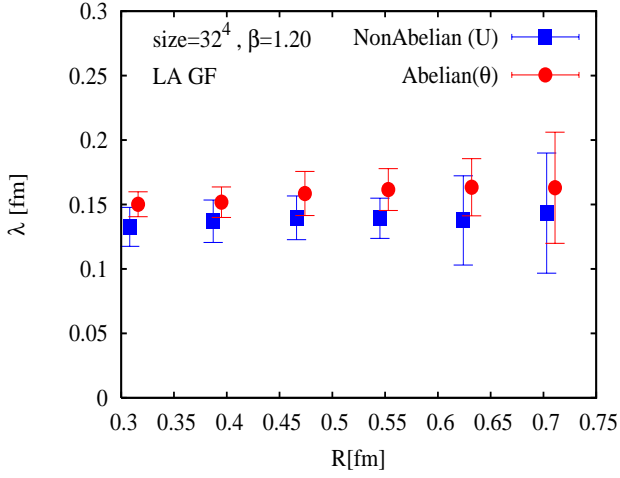}% Here is how to import EPS art
\caption{\label{fig_2} The penetration lengths for the Abelian $\vec{E}_A$ and the non-Abelian $\vec{E}$ electric
field profiles in the Landau gauge and their dependence on the spatial size $R$ of the Wilson loop.}
\end{figure}
\begin{figure}[htb]
\includegraphics[height=6.5cm, width=8.5cm]{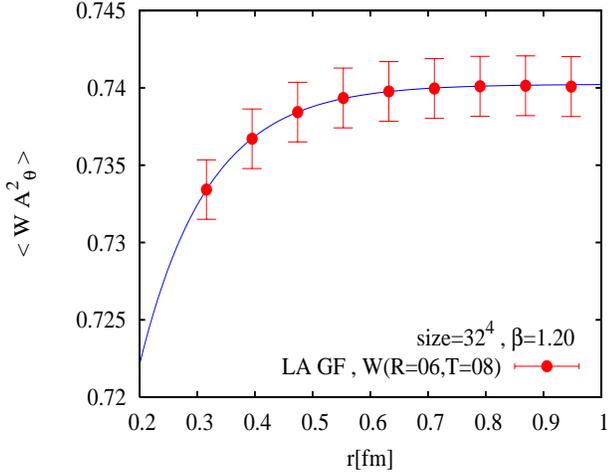}% Here is how to import EPS art
\caption{\label{fig_4} The profiles for the dimension 2 gluon operator in the Landau
gauge around $R\times T= 6\times 8$ Wilson loop. The solid line denotes the best exponential fit by
 the function~\eq{eq:fit} with the best fit parameters given in Table~\ref{tbl:best:fits}.}
\end{figure}
\begin{figure}[htb]
\includegraphics[height=6.5cm, width=8.5cm]{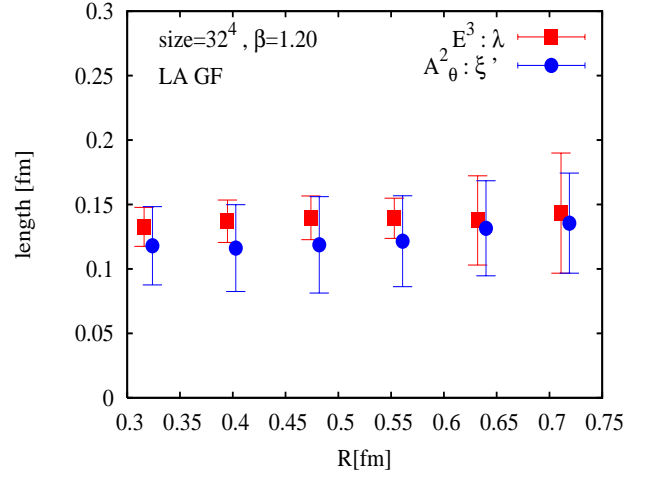}% Here is how to import EPS art
\caption{\label{fig_cp_landau}
The coherence lengths of the dimension 2 gluon operator and the penetration
length in the Landau gauge for various $R$.}
\end{figure}
\begin{figure}[htb]
\includegraphics[height=6.5cm, width=8.5cm]{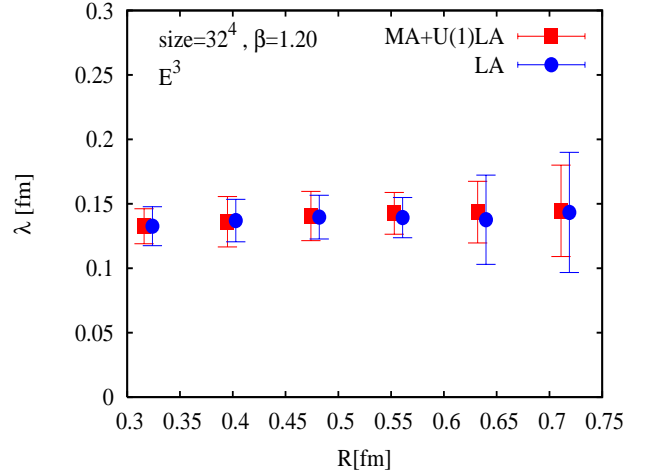}
\caption{\label{comp_penetration}
The penetration lengths of the non-Abelian electric field in the Landau gauge and
in the MA $+$ U1LA gauge for various $R$.}
\end{figure}
\begin{figure}[htb]
\includegraphics[height=6.5cm, width=8.5cm]{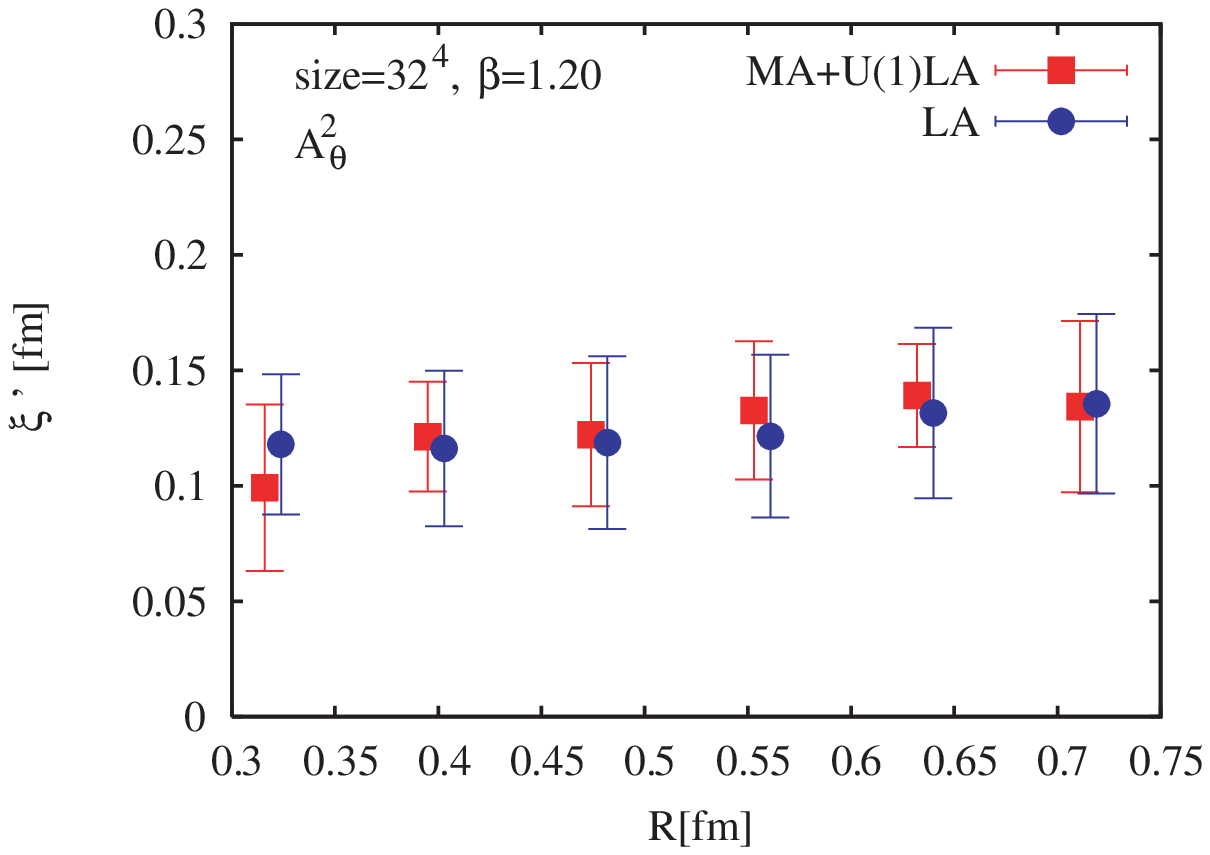}% Here is how to import EPS art
\caption{\label{comp_coherence}
The coherence lengths of the dimension 2 gluon operator in the Landau gauge and in
the MA $+$ U1LA gauge for various $R$.}
\end{figure}

In this gauge, the quantity $A^+A^-(s)$ may have
a physical meaning as we have discussed above. If the remaining $U(1)$ symmetry
is gauge fixed by the $U(1)$ Landau gauge, the dimension 2
gluon operator $A^2(s)$ also acquires a physical meaning.

In order to study different definitions of the quantity
$A^+A^-(s)$, we plot the profile of the $A^+A^-(s)$ condensate in Fig.~\ref{w-AA_theta}
using another definition
\beqn
A^+A^-_\theta = \sum_{s,\mu} \langle\{[\theta^1_\mu(s)]^2+[\theta^2_\mu(s)]^2\}\rangle\,,
\label{eq:AA_theta}
\eeqn
which uses the angles $\theta_\mu(s)$ given by the relation $U_\mu(s)= \exp(i\theta_\mu^a(s)\sigma^a)$.
In Fig.~\ref{w-AAk2_coherence} we show the coherence length determined
by the use of the quantities $A^+A^-_u$, $A^+A^-_\theta$ and $k^2$. From Fig.~\ref{w-AAk2_coherence}, we
conclude that within the error bars these coherence lengths coincide.

It is interesting to determine a non--perturbative content of the gluon operator.
To this end we measure only the monopole contributions to the dimension 2 gluon operator $A^2(s)$
after the MA gauge, and, subsequently, the additional $U(1)$ Landau gauge are fixed.
This way of defining of the non-perturbative quantities is justified because it is known
that in the MA gauge the monopole contributions are responsible for essential non-perturbative physics.
The monopole contribution to the coherence lengths is plotted in Fig.~\ref{fig_6}.

We plot the dependence of the coherence lengths of the spatial extension~$R$ of the Wilson loop
in Fig.~\ref{fig_cp_ma}.
One can clearly see that these values are almost independent on $R$.
Moreover, it is very interesting that the values of the coherence lengths are
almost the same as those of the penetration lengths. Hence, if this relation holds for larger $R$
-- which is very plausible -- then the type of the vacuum must be near the border to the
type 1 and type 2. This is consistent with the results of Ref.~\cite{Cea:1995zt,Singh:1993jj,matsubara94,kato98} obtained
with the help of the classical equations of motions of the dual Ginzburg-Landau theory~\cite{suzuki88}.
However it should be emphasized that in this paper the determination of the vacuum type was done for the first time
quantum-mechanically.

\subsection{The LA gauge case}

The Abelian electric fields $E_{Ai}^a$ are defined in the LA gauge
similarly to the MA gauge. We use the Abelian plaquettes $\theta_{\mu\nu}^a(s)$ defined with the help of
the link variables $\theta_{\mu}^a(s)$:
\begin{eqnarray}
\theta_{\mu\nu}^a(s)\equiv\theta_{\mu}^a(s)+\theta^a_{\nu}(s+\hat{\mu})-\theta_{\mu}^a(
s+\hat{\nu})-\theta^a_{\nu}(s)
\end{eqnarray}
where $\theta_{\mu}^a(s)$ is given by $U_{\mu}(s)=\exp(i\theta_{\mu}^a(s)\sigma^a)$.

The dimension 2 gluonic operator is~\footnote{Note that in Ref.~\cite{fedor2} a different definition of the dimension 2 gluonic
operator was used: $A^2(s) \equiv 2 \Sigma_{\mu=1}^{4} (1-\frac{1}{2}TrU_\mu(s)     )$.
This definition has the same continuum limit as Eq.~(\ref{AS1}).}:
\begin{eqnarray}
A^2(s) \equiv \sum_{\mu=1}^{4} \sum_{a=1}^{3}\left(\theta_\mu^a(s)\right)^2\,.
\label{AS1}
\end{eqnarray}

First let us discuss a determination of the electric fields around a pair of a
static quark and an antiquark in the LA gauge.
Since the confining behavior of the chromoelectric string
is seen for large enough quark-antiquark distances $R$, we
have performed the measurements for various $R$ and $T$.
A typical example is shown in Fig.~\ref{fig_1}.

The $R$ dependence of the penetration lengths is shown in  Fig.~\ref{fig_2}. The maximum quark
distance in Fig.~\ref{fig_2} is $R=0.71$fm which may not be large enough to see the confining string
behavior. On the other hand, we see a small but clear discrepancy between the penetration lengths
of the Abelian $\vec{E}_A$ and the
non-Abelian $\vec{E}$ electric fields. The authors think it is caused by too small distance between the quark
and anti-quark so that the different effects from the Columbic interaction may still play a role.
Anyway for the interquark distances $R\le 0.71$~fm, the Abelian dominance is not observed in the LA gauge.

Now let us discuss the measurements of the coherence lengths.  As it was
explained above, at least in the LA gauge, the operator $A_{\mu}^2$ (or its square-root)
is physically relevant and may have information about properties of a dual Higgs field characterizing the QCD vacuum.
Hence we expect that the coherence length can be measured with the help of $A^2(s)$.
In Fig.~\ref{fig_4} we show a typical example of the $A_{\mu}^2$ profile around the string
where we have adopted the lattice definition (\ref{AS1}) for the operator
$A^2(s)$. This is very exciting, since the behavior of the profile is just what we expect
from a profile of a Higgs field.

We plot the $R$ dependence of the coherence lengths and the penetration
lengths in Fig.~\ref{fig_cp_landau}. It is
very interesting that the values of the coherence lengths are almost the same as those for
the penetration lengths of non-Abelian $\vec{E}$ electric field. Hence
if the observed behavior of the correlations is not changed with the increase of $R$, then
the type of the vacuum should be fixed to be near the border to the type 1 and type 2 as in MA gauge.

\subsection{Comparison  between MA gauge and LA gauge}
In order to consider the gauge-(in)dependence of the
dual superconductor picture, we show in Fig.~\ref{comp_penetration} the penetration length
determined in the MA $+$ U1LA gauge and in the LA gauge.
We also plot the coherence length in Fig.~\ref{comp_coherence}.
{}From these figures, we observe that the coherence and correlation lengths calculated in different gauges coincide
with each other.

\section{Conclusions}

We have observed that
\begin{enumerate}
  \item The coherence lengths of the vacuum of the SU(2) gluodynamics measured in the Maximal Abelian gauge and in the Landau gauge
        are all consistent with each other.
  \item   Since the penetration lengths obtained in the MA gauge are in agreement with the penetration lengths calculated in
          the LA gauge, we conclude that the type of the vacuum in both gauges is near the border between type 1 and type 2.
  \item The monopole contributions to $A^2(s)$ alone reproduces the full coherence length, although the absolute value of the correlations is
   smaller (this last fact is quite natural). Therefore the type of the vacuum can be determined only from the monopole contributions. The
   observed phenomenon is yet another example of the monopole  dominance. It provides an additional support
   to our expectation that the Abelian monopoles are responsible for all non-perturbative phenomena related to the confinement of color in QCD.
\end{enumerate}

\begin{acknowledgments}
The numerical simulations of this work were done using RSCC computer
clusters in RIKEN. The authors would like to thank RIKEN for their
support of computer facilities. T.S. is supported by JSPS Grant-in-Aid
for Scientific Research on Priority Areas 13135210 and (B) 15340073,
M.N.Ch. and M.I.P. are partially supported by RFBR-04-02-16079,
RFBR-05-02-16306a and RFBR-DFG-436-RUS-113/739/0 grants,
and by the EU Integrated Infrastructure Initiative Hadron Physics (I3HP)
under contract RII3-CT-2004-506078.
\end{acknowledgments}


\begin{thebibliography}{99}
\bibitem{tHooft:1975pu}
G.~'t~Hooft, in Proceedings of the EPS International Conference, edited by A.~Zichichi, p. 1225 (1976).
%
\bibitem{Mandelstam:1974pi}
S.~Mandelstam,
%``Vortices And Quark Confinement In Nonabelian Gauge Theories,''
Phys.\ Rept.\  {\bf 23}, 245 (1976).
%%CITATION = PRPLC,23,245;%%
%
\bibitem{tHooft:1981ht}
G.~'t Hooft,
%``Topology Of The Gauge Condition And New Confinement Phases In Nonabelian Gauge Theories,''
Nucl.\ Phys.\ B {\bf 190}, 455 (1981).
%%CITATION = NUPHA,B190,455;%%
%
\bibitem{suzuki-83}
T.~Suzuki, Prog. Theor. Phys. {\bf 69}, 1827(1983).
\bibitem{kronfeld}
A.~S.~Kronfeld, M.~L.~Laursen, G.~Schierholz, U.~J.~Wiese,
%``Monopole Condensation And Color Confinement,''
Phys.\ Lett.\ B {\bf 198}, 516 (1987).
%%CITATION = PHLTA,B198,516;%%
%
\bibitem{AbelianDominance}
T.~Suzuki, I.~Yotsuyanagi, Phys. Rev. {\bf D42}, R4257
(1990);
%%CITATION = PHRVA,D42,4257;%%
\bibitem{Reviews}
T.~Suzuki, Nucl. Phys. Proc. Suppl.  {\bf 30}, 176 (1993);
%%CITATION = NUPHZ,30,176;%%
M.~N.~Chernodub,  M.~I.~Polikarpov,
in "Confinement, duality, and nonperturbative aspects of QCD",
Ed. by P. van Baal, Plenum Press, p. 387 (1997).
%%CITATION = HEP-TH 9710205;%%
\bibitem{bali-96}
G.~S.~Bali, V.~Bornyakov, M.~Muller-Preussker, K.~Schilling,
%``Dual Superconductor Scenario of Confinement: A Systematic Study of Gribov Copy Effects,''
Phys. Rev.  {\bf D54}, 2863 (1996).
%[arXiv:hep-lat/9603012].
%%CITATION = HEP-LAT 9603012;%%
\bibitem{Koma:2003gq}
Y.~Koma, M.~Koma, E.-M.~Ilgenfritz, T.~Suzuki, and M.~I.~Polikarpov,
Phys. Rev. {\bf D68}, 094018 (2003).
 %%CITATION = HEP-LAT 0302006;%%
\bibitem{Singh:1993jj}
V.~Singh D.~A.~Browne, and R.~W.~Haymaker,
 Phys. Lett. {\bf B306}, 115 (1993).
%%CITATION = HEP-LAT 9301004;%%
%
\bibitem{shiba95}
H.~Shiba, T.~Suzuki,
%''Monopole action and condensation in SU(2) QCD'',
Phys. Lett. {\bf B351}, 519 (1995);
%
\bibitem{ref:order:parameter}
M.~N.~Chernodub, M.~I.~Polikarpov and A.~I.~Veselov,
%``Monopole order parameter in SU(2) lattice gauge theory,''
Nucl.\ Phys.\ Proc.\ Suppl.\  {\bf 49}, 307 (1996);
%%CITATION = HEP-LAT 9512030;%
M.~N.~Chernodub, M.~I.~Polikarpov and A.~I.~Veselov,
%``Effective constraint potential for Abelian monopole in SU(2) lattice  gauge theory,''
Phys.\ Lett.\ B {\bf 399}, 267 (1997);
%%CITATION = HEP-LAT 9610007;%%
A.~Di Giacomo and G.~Paffuti,
%``A disorder parameter for dual superconductivity in gauge theories,''
Phys.\ Rev.\ D {\bf 56}, 6816 (1997).
%%CITATION = HEP-LAT 9707003;%%

\bibitem{Cea:1995zt}
P.~Cea, L.~Cosmai,
 Phys. Rev. {\bf D52}, 5152 (1995).
\bibitem{matsubara94}
Y.~Matsubara, S.~Ejiri, T.~Suzuki,
%''The (dual) Meissner effect in SU(2) and SU(3) QCD''
Nucl. Phys. Proc. Suppl. {\bf 34}, 176 (1994).
\bibitem{kato98}
S.~Kato, M.~N.~Chernodub, S.~Kitahara, N.~Nakamura, M.~I.~Polikarpov, T.~Suzuki,
%''Various representations of infrared effective lattice QCD''
Nucl. Phys. Proc. Suppl. {\bf 63}, 471 (1998).
\bibitem{bali98}
G.~S.~Bali, C.~Schlichter, K.~Schilling,
%''Probing the QCD Vacuum with Static Sources in Maximal Abelian Projection''
Prog. Theor. Phys. Suppl. {\bf 131}, 645 (1998).
\bibitem{suzuki88}
T.~Suzuki,
Prog. Theor. Phys. {\bf 80}, 929 (1988).
\bibitem{maedan89}
S.~Maedan, T.~Suzuki,
Prog. Theor. Phys. {\bf 81}, 229 (1989).
\bibitem{screening}
T.~Suzuki, M.N.~Chernodub, Phys. Lett. {\bf B563}, 183 (2003).
\bibitem{suzuki05}
T.~Suzuki, K.~Ishiguro, Y.~Mori, T.~Sekido,
%''The dual Meissner effect and Abelian magnetic displacement currents''
%hep-lat/0410001;
Phys. Rev. Lett. {\bf 94}, 132001 (2005);
%''The dual Meissner effect in SU(2) Landau gauge''
hep-lat/0410039.
\bibitem{DeGrand:1980eq}
T.~A.~DeGrand, D.~Toussaint,
 Phys. Rev. {\bf D22}, 2478 (1980).
\bibitem{fedor02}
F.~V.~Gubarev, V.~I.~Zakharov,
{\it ''Gauge invariant monopoles in SU(2) gluodynamics''}, hep-lat/0204017;
F.~V.~Gubarev,
{\it ''Gauge invariant monopoles in lattice SU(2) gluodynamics''}, hep-lat/0204018.
\bibitem{fedor05}
F.~V.~Gubarev, S.~M.~Morozov,
{\it ''$\langle A^2 \rangle$ Condensate, Bianchi Identities and
Chromomagnetic Fields Degeneracy in SU(2) YM Theory''}, hep-lat/0503023;
\bibitem{maxim05}
M.~N.~Chernodub,
{\it ''A gauge-invariant object in non-Abelian gauge theory''}, hep-lat/0503018;
\bibitem{Chernodub:2005jh}
M.~N.~Chernodub,
{\it ``Yang-Mills theory in Landau gauge as a liquid crystal''},
hep-th/0506107;
%%CITATION = HEP-TH 0506107;%%
{\it ``Liquid crystal defects and confinement in Yang-Mills theory''},
hep-th/0507221.
%%CITATION = HEP-TH 0507221;%%
\bibitem{fedor1}
F.~V.~Gubarev, L.~Stodolsky, V.~I.~Zakharov,
Phys.Rev.Lett. {\bf 86}, 2220 (2001).
\bibitem{compactQED}
A.~M.~Polyakov, Phys. Lett. {\bf 59B}, 82 (1975);
M.~E.~Peshkin, Ann. Phys. {\bf 113}, 122 (1978).
\bibitem{fedor2}
F.~V.~Gubarev, V.~I.~Zakharov,
%''On the Emerging Phenomenology of <(A_\mu)^2>''
Phys. Lett. {\bf 501}, 28 (2001).
\bibitem{stodolsky}
L.~Stodolsky, Pierre van Baal, V.~I.~Zakharov,
Phys. Lett. {\bf B552}, 214 (2003) and references therein.
\bibitem{Mass}
M.~J.~Lavelle, M.~Schaden, Phys. Lett. {\bf B208}, 297 (1988);
M.~Lavelle, M.~Oleszczuk, Mod. Phys. Lett. {\bf A7}, 3617 (1992);
D.~Dudal et al., JHEP {\bf 0401}, 044 (2004);
P.~Boucaud et al., Phys. Rev. {\bf D66}, 034504 (2002);
X.~Li, C.~M.~Shakin, "The Gluon Propagator in Minkowski and Euclidean Space: Role of
an $A^2$ Condensate", hep-ph/0411234 and references therein.
%
\bibitem{kondo}
K-I.~Kondo, Phys. Lett. {\bf B514}, 335 (2001);
Phys.Lett. {\bf B572}, 210 (2003).
%hep-th/0306195
\bibitem{ref:string:MA:gauge}
%G.~S.~Bali, C.~Schlichter, K.~Schilling,
%{\it ``Probing the QCD vacuum with static sources in maximal Abelian projection''},
%Prog.\ Theor.\ Phys.\ Suppl.\  {\bf 131}, 645 (1998);
%%CITATION = HEP-LAT 9802005;%%
F.~V.~Gubarev, E.~M.~Ilgenfritz, M.~I.~Polikarpov, T.~Suzuki,
%{\it ``The lattice SU(2) confining string as an Abrikosov vortex''},
Phys.\ Lett.\ B {\bf 468}, 134 (1999);
%%CITATION = HEP-LAT 9909099;%%
Y.~Koma, M.~Koma, E.~M.~Ilgenfritz, T.~Suzuki,
%{\it ``A detailed study of the Abelian-projected SU(2) flux tube and its dual Ginzburg-Landau analysis'',}
Phys.\ Rev.\ D {\bf 68}, 114504 (2003).
%%CITATION = HEP-LAT 0308008;%%

\bibitem{ref:Bettencourt}
L.~M.~A.~Bettencourt, R.~J.~Rivers,
%``Interactions between U(1) cosmic strings: An Analytical study,''
Phys.\ Rev.\ D {\bf 51}, 1842 (1995).
%%CITATION = HEP-PH 9405222;%%

\bibitem{ref:Abrikosov}
A.~A.~Abrikosov,
%``On The Magnetic Properties Of Superconductors Of The Second Group,''
Sov.\ Phys.\ JETP {\bf 5}, 1174 (1957)
[Zh.\ Eksp.\ Teor.\ Fiz.\  {\bf 32}, 1442 (1957)].
%%CITATION = SPHJA,5,1174;%%
%
\bibitem{ref:NO}
H.~B.~Nielsen, P.~Olesen,
%``Vortex-Line Models For Dual Strings,''
Nucl.\ Phys.\ B {\bf 61}, 45 (1973).
%%CITATION = NUPHA,B61,45;%%

\bibitem{ref:Hart}
A.~Hart, M.~Teper,
%``Monopole clusters in Abelian projected gauge theories,''
Phys.\ Rev.\ D {\bf 58}, 014504 (1998).
%%CITATION = HEP-LAT 9712003;%%

\bibitem{ref:Zakharov}
V.~G.~Bornyakov, P.~Y.~Boyko, M.~I.~Polikarpov, V.~I.~Zakharov,
%{\it ``Monopole clusters at short and large distances''}
Nucl.\ Phys.\ B {\bf 672}, 222 (2003);
%%CITATION = HEP-LAT 0305021;%%
%MCH added reference
M.~N.~Chernodub, V.~I.~Zakharov,
%``Towards understanding structure of the monopole clusters,''
Nucl.\ Phys.\ B {\bf 669}, 233 (2003)
%%CITATION = HEP-TH 0211267;%%
see also discussion in V.~I.~Zakharov,
{\it ``Dual string from lattice Yang-Mills theory''}
hep-ph/0501011.
%%CITATION = HEP-PH 0501011;%%

\bibitem{ref:monopole:density}
V.~G.~Bornyakov {\it et al.}  [DIK Collaboration],
%{\it ``Dynamics of monopoles and flux tubes in two-flavor dynamical QCD''},
Phys.\ Rev.\ D {\bf 70}, 074511 (2004).
%%CITATION = HEP-LAT 0310011;%%

\bibitem{ref:Akhmedov}
See, for example,
P.~Orland,
%``Extrinsic curvature dependence of Nielsen-Olesen strings,''
Nucl.\ Phys.\ B {\bf 428}, 221 (1994);
%%CITATION = HEP-TH 9404140;%%
E.~T.~Akhmedov, M.~N.~Chernodub, M.~I.~Polikarpov, M.~A.~Zubkov,
%{\it ``Quantum theory of strings in Abelian Higgs model''},
Phys.\ Rev.\ D {\bf 53}, 2087 (1996), and references therein.
%%CITATION = HEP-TH 9505070;%%

\bibitem{iwasaki}
Y.~Iwasaki, Nucl. Phys. {\bf B258}, 141 (1985); Univ. Tsukuba Preprint UTHEP-118(1983) unpublished.
\bibitem{APE}
APE Collaboration: M.~Albanese et al., Phys. Lett. {\bf B192} 163 (1987).
\bibitem{bali-94}
G.~S.~Bali, K.~Schilling, Ch.~Schlichter, Phys. Rev. {\bf D51}, 5165 (1995).
%%CITATION = PHRVA,D22,2478;%%

\bibitem{ref:ambiguity}
S.~Hioki, S.~Kitahara, Y.~Matsubara, O.~Miyamura, S.~Ohno, T.~Suzuki,
%{\it ``Gauge fixing ambiguity and monopole''},
Phys.\ Lett.\ B {\bf 271}, 201 (1991).
%%CITATION = PHLTA,B271,201;%%
%
\bibitem{ref:anatomy}
V.~G.~Bornyakov, M.~N.~Chernodub, F.~V.~Gubarev, M.~I.~Polikarpov, T.~Suzuki, A.~I.~Veselov, V.~I.~Zakharov,
%{\it ``Anatomy of the lattice magnetic monopoles''},
Phys.\ Lett.\ B {\bf 537}, 291 (2002).
%%CITATION = HEP-LAT 0103032;%%

\end{thebibliography}
\end{document}